\documentclass[12pt]{article}

\pdfoutput=1

\usepackage{jheppub}
\usepackage{amsmath,amssymb,amsfonts,amscd,mathrsfs}
\usepackage{xcolor}
\definecolor{darkblue}{rgb}{0.1,0.1,.7}

\usepackage{tablefootnote,colortbl}

\usepackage{braket}

\usepackage{cancel}

\usepackage{arydshln}
\usepackage{booktabs,multirow}

\usepackage{tikz-feynman}
 \tikzfeynmanset{compat=1.0.0}
 
\newcommand\myline{%
	\,\tikz[baseline]\draw[ thick,dashed](0,-\dp\strutbox)--(0,\ht\strutbox);\,%
}

\definecolor{myorange}{RGB}{199,146,32}

\definecolor{Gray1}{gray}{0.97}
\definecolor{Gray2}{gray}{0.9}
\definecolor{LightCyan}{rgb}{0.88,1,1}
\definecolor{blu}{rgb}{0,0,1}

\usepackage{ragged2e}
\usepackage{array}
\newcolumntype{L}[1]{>{\raggedright\let\newline\\\arraybackslash\hspace{0pt}}m{#1}}
\newcolumntype{C}[1]{>{\centering\let\newline\\\arraybackslash\hspace{0pt}}m{#1}}
\newcolumntype{R}[1]{>{\raggedleft\let\newline\\\arraybackslash\hspace{0pt}}m{#1}}

\usepackage{dsfont} 
\usepackage{tcolorbox}
\usepackage{booktabs}

\usepackage{titlesec}
\titleformat*{\section}{\large\bfseries}
\titleformat*{\subsection}{\normalsize\bfseries}
\titleformat*{\subsubsection}{\normalsize\it}
\titleformat*{\paragraph}{\normalsize\bfseries}
\titleformat*{\subparagraph}{\normalsize\bfseries}

\def\D {{\Delta_T}} 
\def\Dp {{\Delta_T^\prime}}

\newcommand{\reef}[1]{(\ref{#1})}

\def\eps{\epsilon}
\def\Eps{{\cal E}}

\newcommand{\beq}{\begin{equation}} 
\newcommand{\eeq}{\end{equation}}

\def\calO {{\cal O}}

\def\calH {{\cal H}}

\def\calA {{\cal A}} 
 
\def\calE {{\cal E}}

\def\ge{\geqslant}
\def\le{\leqslant}
\def\geq{\geqslant}
\def\leq{\leqslant}
\newcommand{\diffop}[2]{\ifthenelse{\equal{#2}{1}}{\frac{\mrm{d}}{\mrm{d} #1}}{\frac{\mrm{d}^#2}{\mrm{d} #1^#2}}}

\def\L{\langle}
\def\R{\rangle}

\newcommand{\mrm}[1]{{\mathrm #1}}

\usepackage[normalem]{ulem}

\def \H{\mathcal{H}}

\newcommand{\roig}{\color{red}}

\newcommand{\be}{\begin{equation}}
\newcommand{\ee}{\end{equation}}
\def\bea#1\eea{\begin{align}#1\end{align}}

  \def\th{\theta}

\newcommand{\pd}[2]{\frac{\partial #1}{\partial #2}} 
\newcommand{\matrixel}[3]{\left< #1 \vphantom{#2#3} \right|
	#2 \left| #3 \vphantom{#1#2} \right>} 

\usepackage{accents}
\newlength{\dhatheight}

\numberwithin{equation}{section}
\usepackage{tocloft}
    \newcommand{\dVdos}{
 \begin{minipage}[h]{0.085\linewidth}
\begin{tikzpicture}
\begin{feynman}[small]
 \node [dot] (i1) at (0,.35);
 \node [dot] (i2) at (.8,.35);  
 \node [dot] (j1) at (.4,.9);
 \node [dot] (j2) at (1.2,.9);  
   \vertex (A1) at (0.2,1.3); 
     \vertex (B1) at (0.2,-.15) ; 
     \vertex (B1t) at (0.2,-.33)  {{\roig \scriptsize{$^{E_1}$}}}; 
   \vertex (A2) at (.6,1.5); 
    \vertex (A2t) at (.6,1.6) {{\roig \scriptsize{$^{E_1+E_2}$}}}; 
     \vertex (B2) at (.6,-.15); 
   \vertex (A3) at (1,1.3); 
     \vertex (B3) at (1,-.15); 
     \vertex (B3t) at (1,-.33) {{\roig \scriptsize{$^{E_2}$}}}; 
\diagram*{
   (i1) -- [ quarter right, looseness=.3,  thick] (i2)  -- [ quarter right, looseness=.3 ,  thick] (i1) ,
     (i1) -- [ half right, looseness=.8,  thick] (i2)  -- [ half right, looseness=.8 ,  thick] (i1) ,
   (j1) -- [ quarter right, looseness=.3,  thick] (j2)  -- [ quarter right, looseness=.3 ,  thick] (j1) ,
     (j1) -- [ half right, looseness=.8,  thick] (j2)  -- [ half right, looseness=.8 ,  thick] (j1) ,
       (A1) -- [scalar, thick, red] (B1) ,
        (A2) -- [scalar, thick, red] (B2) ,
        (A3) -- [scalar, thick, red] (B3) 
};
  \end{feynman}
\end{tikzpicture}
  \end{minipage}  
  }
  
    \newcommand{\dVtres}{
 \begin{minipage}[h]{0.085\linewidth}
\begin{tikzpicture}
\begin{feynman}[small]
 \node [dot] (i1) at (0,.35);
 \node [dot] (i2) at (1.2,.35);  
 \node [dot] (j1) at (.4,.9);
 \node [dot] (j2) at (.8,.9);  
   \vertex (A1) at (0.2,1.3); 
     \vertex (B1) at (0.2,-.15) ; 
     \vertex (B1t) at (0.2,-.33)  {{\roig \scriptsize{$^{E_1}$}}}; 
   \vertex (A2) at (.6,1.5); 
    \vertex (A2t) at (.6,1.6) {{\roig \scriptsize{$^{E_1+E_2}$}}}; 
     \vertex (B2) at (.6,-.15); 
   \vertex (A3) at (1,1.3); 
     \vertex (B3) at (1,-.15); 
     \vertex (B3t) at (1,-.33) {{\roig \scriptsize{$^{E_2}$}}}; 
\diagram*{
   (i1) -- [ quarter right, looseness=.3,  thick] (i2)  -- [ quarter right, looseness=.3 ,  thick] (i1) ,
     (i1) -- [ half right, looseness=.6,  thick] (i2)  -- [ half right, looseness=.6 ,  thick] (i1) ,
   (j1) -- [ quarter right, looseness=.5,  thick] (j2)  -- [ quarter right, looseness=.5 ,  thick] (j1) ,
     (j1) -- [ half right, looseness=1.2,  thick] (j2)  -- [ half right, looseness=1.28 ,  thick] (j1) ,
       (A1) -- [scalar, thick, red] (B1) ,
        (A2) -- [scalar, thick, red] (B2) ,
        (A3) -- [scalar, thick, red] (B3) 
};
  \end{feynman}
\end{tikzpicture}
  \end{minipage}  
  }
  
    \newcommand{\dun}{
 \begin{minipage}[h]{0.1\linewidth}
\begin{tikzpicture}
\begin{feynman}[small]
 \node [dot] (I1) at (-.15,.85);
 \node [dot] (J1) at (.45,.85);  
 \node [dot] (I2) at (.15,0.35);
 \node [dot] (J2) at (1.05,0.35);  
  \node [dot] (I3) at (.75,.85);
 \node [dot] (J3) at (1.35,.85);  
   \vertex (A1) at (0,1.2); 
     \vertex (B1) at (0,0); 
   \vertex (A2) at (+.3,1.2); 
     \vertex (B2) at (+.3,0); 
   \vertex (A3) at (+2*.3,1.2); 
     \vertex (B3) at (+2*.3,0); 
   \vertex (A4) at (+3*.3,1.2); 
     \vertex (B4) at (3*.3,0); 
   \vertex (A5) at (4*.3,1.2); 
     \vertex (B5) at (4*.3,0); 
\diagram*{
   (I1) -- [ quarter right, looseness=.3,  thick] (J1)  -- [ quarter right, looseness=.3 ,  thick] (I1) ,
     (I1) -- [ half right, looseness=.8,  thick] (J1)  -- [ half right, looseness=.8 ,  thick] (I1) ,
   (I2) -- [ quarter right, looseness=.3,  thick] (J2)  -- [ quarter right, looseness=.3 ,  thick] (I2) ,
     (I2) -- [ half right, looseness=.8,  thick] (J2)  -- [ half right, looseness=.8 ,  thick] (I2) ,
   (I3) -- [ quarter right, looseness=.3,  thick] (J3)  -- [ quarter right, looseness=.3 ,  thick] (I3) ,
     (I3) -- [ half right, looseness=.8,  thick] (J3)  -- [ half right, looseness=.8 ,  thick] (I3) ,
        (A1) -- [scalar, thick, red] (B1) ,
        (A2) -- [scalar, thick, red] (B2) ,
       (A3) -- [scalar, thick, red] (B3) ,
       (A4) -- [scalar, thick, red] (B4) ,
       (A5) -- [scalar, thick, red] (B5) ,
};
  \end{feynman}
\end{tikzpicture}
  \end{minipage}  
  }
  
    \newcommand{\ddos}{
 \begin{minipage}[h]{0.1\linewidth}
\begin{tikzpicture}
\begin{feynman}[small]
 \node [dot] (I1) at (.15,.85);
 \node [dot] (J1) at (.45,.85);  
 \node [dot] (I2) at (-.15,0.35);
 \node [dot] (J2) at (1.05,0.35);  
  \node [dot] (I3) at (.75,.85);
 \node [dot] (J3) at (1.35,.85);  
   \vertex (A1) at (0,1.2); 
     \vertex (B1) at (0,0); 
   \vertex (A2) at (+.3,1.2); 
     \vertex (B2) at (+.3,0); 
   \vertex (A3) at (+2*.3,1.2); 
     \vertex (B3) at (+2*.3,0); 
   \vertex (A4) at (+3*.3,1.2); 
     \vertex (B4) at (3*.3,0); 
   \vertex (A5) at (4*.3,1.2); 
     \vertex (B5) at (4*.3,0); 
\diagram*{
   (I1) -- [ quarter right, looseness=.5,  thick] (J1)  -- [ quarter right, looseness=.5 ,  thick] (I1) ,
     (I1) -- [ half right, looseness=1.2,  thick] (J1)  -- [ half right, looseness=1.2 ,  thick] (I1) ,
   (I2) -- [ quarter right, looseness=.25,  thick] (J2)  -- [ quarter right, looseness=.25 ,  thick] (I2) ,
     (I2) -- [ half right, looseness=.55,  thick] (J2)  -- [ half right, looseness=.55 ,  thick] (I2) ,
   (I3) -- [ quarter right, looseness=.3,  thick] (J3)  -- [ quarter right, looseness=.3 ,  thick] (I3) ,
     (I3) -- [ half right, looseness=.8,  thick] (J3)  -- [ half right, looseness=.8 ,  thick] (I3) ,
        (A1) -- [scalar, thick, red] (B1) ,
        (A2) -- [scalar, thick, red] (B2) ,
       (A3) -- [scalar, thick, red] (B3) ,
       (A4) -- [scalar, thick, red] (B4) ,
       (A5) -- [scalar, thick, red] (B5) ,
};
  \end{feynman}
\end{tikzpicture}
  \end{minipage}  
  }

    \newcommand{\dtres}{
 \begin{minipage}[h]{0.1\linewidth}
\begin{tikzpicture}
\begin{feynman}[small]
 \node [dot] (I1) at (.75,.85);
 \node [dot] (J1) at (1.05,.85);  
 \node [dot] (I2) at (.15,0.35);
 \node [dot] (J2) at (1.35,0.35);  
  \node [dot] (I3) at (-.15,.85);
 \node [dot] (J3) at (.45,.85);  
   \vertex (A1) at (0,1.2); 
     \vertex (B1) at (0,0); 
   \vertex (A2) at (+.3,1.2); 
     \vertex (B2) at (+.3,0); 
   \vertex (A3) at (+2*.3,1.2); 
     \vertex (B3) at (+2*.3,0); 
   \vertex (A4) at (+3*.3,1.2); 
     \vertex (B4) at (3*.3,0); 
   \vertex (A5) at (4*.3,1.2); 
     \vertex (B5) at (4*.3,0); 
\diagram*{
   (I1) -- [ quarter right, looseness=.5,  thick] (J1)  -- [ quarter right, looseness=.5 ,  thick] (I1) ,
     (I1) -- [ half right, looseness=1.2,  thick] (J1)  -- [ half right, looseness=1.2 ,  thick] (I1) ,
   (I2) -- [ quarter right, looseness=.25,  thick] (J2)  -- [ quarter right, looseness=.25 ,  thick] (I2) ,
     (I2) -- [ half right, looseness=.55,  thick] (J2)  -- [ half right, looseness=.55 ,  thick] (I2) ,
   (I3) -- [ quarter right, looseness=.3,  thick] (J3)  -- [ quarter right, looseness=.3 ,  thick] (I3) ,
     (I3) -- [ half right, looseness=.8,  thick] (J3)  -- [ half right, looseness=.8 ,  thick] (I3) ,
        (A1) -- [scalar, thick, red] (B1) ,
        (A2) -- [scalar, thick, red] (B2) ,
       (A3) -- [scalar, thick, red] (B3) ,
       (A4) -- [scalar, thick, red] (B4) ,
       (A5) -- [scalar, thick, red] (B5) ,
};
  \end{feynman}
\end{tikzpicture}
  \end{minipage}  
  }

    \newcommand{\dquatre}{
 \begin{minipage}[h]{0.1\linewidth}
\begin{tikzpicture}
\begin{feynman}[small]
 \node [dot] (I1) at (.75,.85);
 \node [dot] (J1) at (1.05,.85);  
 \node [dot] (I2) at (-.15,0.35);
 \node [dot] (J2) at (1.35,0.35);  
  \node [dot] (I3) at (.15,.85);
 \node [dot] (J3) at (.45,.85);  
   \vertex (A1) at (0,1.15); 
     \vertex (B1) at (0,0); 
   \vertex (A2) at (+.3,1.15); 
     \vertex (B2) at (+.3,0); 
   \vertex (A3) at (+2*.3,1.15); 
     \vertex (B3) at (+2*.3,0); 
   \vertex (A4) at (+3*.3,1.15); 
     \vertex (B4) at (3*.3,0); 
   \vertex (A5) at (4*.3,1.15); 
     \vertex (B5) at (4*.3,0); 
\diagram*{
   (I1) -- [ quarter right, looseness=.5,  thick] (J1)  -- [ quarter right, looseness=.5 ,  thick] (I1) ,
     (I1) -- [ half right, looseness=1.2,  thick] (J1)  -- [ half right, looseness=1.2 ,  thick] (I1) ,
   (I2) -- [ quarter right, looseness=.2,  thick] (J2)  -- [ quarter right, looseness=.2 ,  thick] (I2) ,
     (I2) -- [ half right, looseness=.45,  thick] (J2)  -- [ half right, looseness=.45 ,  thick] (I2) ,
   (I3) -- [ quarter right, looseness=.5,  thick] (J3)  -- [ quarter right, looseness=.5,  thick] (I3) ,
     (I3) -- [ half right, looseness=1.2,  thick] (J3)  -- [ half right, looseness=1.2 ,  thick] (I3) ,
        (A1) -- [scalar, thick, red] (B1) ,
        (A2) -- [scalar, thick, red] (B2) ,
       (A3) -- [scalar, thick, red] (B3) ,
       (A4) -- [scalar, thick, red] (B4) ,
       (A5) -- [scalar, thick, red] (B5) ,
};
  \end{feynman}
\end{tikzpicture}
  \end{minipage}  
  }

    \newcommand{\dcinc}{
 \begin{minipage}[h]{0.1\linewidth}
\begin{tikzpicture}
\begin{feynman}[small]
 \node [dot] (I1) at (-.15,1.47);
 \node [dot] (J1) at (.75,1.47);  
 \node [dot] (I2) at (.15,0.9);
 \node [dot] (J2) at (1.05,0.9);   
  \node [dot] (I3) at (.45,.33);
 \node [dot] (J3) at (1.35,.33); 
   \vertex (A1) at (0,1.8); 
     \vertex (B1) at (0,0); 
   \vertex (A2) at (+.3,1.8); 
     \vertex (B2) at (+.3,0); 
   \vertex (A3) at (+2*.3,1.8); 
     \vertex (B3) at (+2*.3,0); 
   \vertex (A4) at (+3*.3,1.8); 
     \vertex (B4) at (3*.3,0); 
   \vertex (A5) at (4*.3,1.8); 
     \vertex (B5) at (4*.3,0); 
\diagram*{
      (I1) -- [ quarter right, looseness=.3,  thick] (J1)  -- [ quarter right, looseness=.3 ,  thick] (I1) ,
     (I1) -- [ half right, looseness=.8,  thick] (J1)  -- [ half right, looseness=.8 ,  thick] (I1) ,
   (I2) -- [ quarter right, looseness=.3,  thick] (J2)  -- [ quarter right, looseness=.3 ,  thick] (I2) ,
     (I2) -- [ half right, looseness=.8,  thick] (J2)  -- [ half right, looseness=.8 ,  thick] (I2) ,
   (I3) -- [ quarter right, looseness=.3,  thick] (J3)  -- [ quarter right, looseness=.3 ,  thick] (I3) ,
     (I3) -- [ half right, looseness=.8,  thick] (J3)  -- [ half right, looseness=.8 ,  thick] (I3) ,
        (A1) -- [scalar, thick, red] (B1) ,
        (A2) -- [scalar, thick, red] (B2) ,
       (A3) -- [scalar, thick, red] (B3) ,
       (A4) -- [scalar, thick, red] (B4) ,
       (A5) -- [scalar, thick, red] (B5) ,
};
  \end{feynman}
\end{tikzpicture}
  \end{minipage}  
  }

    \newcommand{\dsis}{
 \begin{minipage}[h]{0.1\linewidth}
\begin{tikzpicture}
\begin{feynman}[small]
 \node [dot] (I1) at (.45,1.47);
 \node [dot] (J1) at (1.35,1.47);  
 \node [dot] (I2) at (-.15,0.9);
 \node [dot] (J2) at (1.05,0.9);   
  \node [dot] (I3) at (.15,.33);
 \node [dot] (J3) at (.75,.33); 
   \vertex (A1) at (0,1.8); 
     \vertex (B1) at (0,0); 
   \vertex (A2) at (+.3,1.8); 
     \vertex (B2) at (+.3,0); 
   \vertex (A3) at (+2*.3,1.8); 
     \vertex (B3) at (+2*.3,0); 
   \vertex (A4) at (+3*.3,1.8); 
     \vertex (B4) at (3*.3,0); 
   \vertex (A5) at (4*.3,1.8); 
     \vertex (B5) at (4*.3,0); 
\diagram*{
      (I1) -- [ quarter right, looseness=.3,  thick] (J1)  -- [ quarter right, looseness=.3 ,  thick] (I1) ,
     (I1) -- [ half right, looseness=.8,  thick] (J1)  -- [ half right, looseness=.8 ,  thick] (I1) ,
   (I2) -- [ quarter right, looseness=.25,  thick] (J2)  -- [ quarter right, looseness=.25 ,  thick] (I2) ,
     (I2) -- [ half right, looseness=.55,  thick] (J2)  -- [ half right, looseness=.55 ,  thick] (I2) ,
   (I3) -- [ quarter right, looseness=.3,  thick] (J3)  -- [ quarter right, looseness=.3 ,  thick] (I3) ,
     (I3) -- [ half right, looseness=.8,  thick] (J3)  -- [ half right, looseness=.8 ,  thick] (I3) ,
        (A1) -- [scalar, thick, red] (B1) ,
        (A2) -- [scalar, thick, red] (B2) ,
       (A3) -- [scalar, thick, red] (B3) ,
       (A4) -- [scalar, thick, red] (B4) ,
       (A5) -- [scalar, thick, red] (B5) ,
};
  \end{feynman}
\end{tikzpicture}
  \end{minipage}  
  }

    \newcommand{\dsishc}{
 \begin{minipage}[h]{0.1\linewidth}
\begin{tikzpicture}
\begin{feynman}[small]
 \node [dot] (I1) at (-.15,1.47);
 \node [dot] (J1) at (.75,1.47);  
 \node [dot] (I2) at (.15,0.9);
 \node [dot] (J2) at (1.35,0.9);   
  \node [dot] (I3) at (.45,.33);
 \node [dot] (J3) at (1.05,.33); 
   \vertex (A1) at (0,1.8); 
     \vertex (B1) at (0,0); 
   \vertex (A2) at (+.3,1.8); 
     \vertex (B2) at (+.3,0); 
   \vertex (A3) at (+2*.3,1.8); 
     \vertex (B3) at (+2*.3,0); 
   \vertex (A4) at (+3*.3,1.8); 
     \vertex (B4) at (3*.3,0); 
   \vertex (A5) at (4*.3,1.8); 
     \vertex (B5) at (4*.3,0); 
\diagram*{
      (I1) -- [ quarter right, looseness=.3,  thick] (J1)  -- [ quarter right, looseness=.3 ,  thick] (I1) ,
     (I1) -- [ half right, looseness=.8,  thick] (J1)  -- [ half right, looseness=.8 ,  thick] (I1) ,
   (I2) -- [ quarter right, looseness=.25,  thick] (J2)  -- [ quarter right, looseness=.25 ,  thick] (I2) ,
     (I2) -- [ half right, looseness=.55,  thick] (J2)  -- [ half right, looseness=.55 ,  thick] (I2) ,
   (I3) -- [ quarter right, looseness=.3,  thick] (J3)  -- [ quarter right, looseness=.3 ,  thick] (I3) ,
     (I3) -- [ half right, looseness=.8,  thick] (J3)  -- [ half right, looseness=.8 ,  thick] (I3) ,
        (A1) -- [scalar, thick, red] (B1) ,
        (A2) -- [scalar, thick, red] (B2) ,
       (A3) -- [scalar, thick, red] (B3) ,
       (A4) -- [scalar, thick, red] (B4) ,
       (A5) -- [scalar, thick, red] (B5) ,
};
  \end{feynman}
\end{tikzpicture}
  \end{minipage}  
  }

    \newcommand{\dset}{
 \begin{minipage}[h]{0.1\linewidth}
\begin{tikzpicture}
\begin{feynman}[small]
 \node [dot] (I1) at (.45,1.47);
 \node [dot] (J1) at (.75,1.47);  
 \node [dot] (I2) at (-.15,0.9);
 \node [dot] (J2) at (1.05,0.9);   
  \node [dot] (I3) at (.15,.33);
 \node [dot] (J3) at (1.35,.33); 
   \vertex (A1) at (0,1.8); 
     \vertex (B1) at (0,0); 
   \vertex (A2) at (+.3,1.8); 
     \vertex (B2) at (+.3,0); 
   \vertex (A3) at (+2*.3,1.8); 
     \vertex (B3) at (+2*.3,0); 
   \vertex (A4) at (+3*.3,1.8); 
     \vertex (B4) at (3*.3,0); 
   \vertex (A5) at (4*.3,1.8); 
     \vertex (B5) at (4*.3,0); 
\diagram*{
   (I1) -- [ quarter right, looseness=.5,  thick] (J1)  -- [ quarter right, looseness=.5 ,  thick] (I1) ,
     (I1) -- [ half right, looseness=1.2,  thick] (J1)  -- [ half right, looseness=1.2 ,  thick] (I1) ,
   (I2) -- [ quarter right, looseness=.25,  thick] (J2)  -- [ quarter right, looseness=.25 ,  thick] (I2) ,
     (I2) -- [ half right, looseness=.55,  thick] (J2)  -- [ half right, looseness=.55 ,  thick] (I2) ,
   (I3) -- [ quarter right, looseness=.25,  thick] (J3)  -- [ quarter right, looseness=.25 ,  thick] (I3) ,
     (I3) -- [ half right, looseness=.55,  thick] (J3)  -- [ half right, looseness=.55 ,  thick] (I3) ,
        (A1) -- [scalar, thick, red] (B1) ,
        (A2) -- [scalar, thick, red] (B2) ,
       (A3) -- [scalar, thick, red] (B3) ,
       (A4) -- [scalar, thick, red] (B4) ,
       (A5) -- [scalar, thick, red] (B5) ,
};
  \end{feynman}
\end{tikzpicture}
  \end{minipage}  
  }

    \newcommand{\dvuit}{
 \begin{minipage}[h]{0.1\linewidth}
\begin{tikzpicture}
\begin{feynman}[small]
 \node [dot] (I1) at (.45,1.49);
 \node [dot] (J1) at (1.05,1.49);  
 \node [dot] (I2) at (.15,1.);
 \node [dot] (J2) at (.75,1.);   
  \node [dot] (I3) at (-.15,.39);
 \node [dot] (J3) at (1.35,.39); 
   \vertex (A1) at (0,1.8); 
     \vertex (B1) at (0,0); 
   \vertex (A2) at (+.3,1.8); 
     \vertex (B2) at (+.3,0); 
   \vertex (A3) at (+2*.3,1.8); 
     \vertex (B3) at (+2*.3,0); 
   \vertex (A4) at (+3*.3,1.8); 
     \vertex (B4) at (3*.3,0); 
   \vertex (A5) at (4*.3,1.8); 
     \vertex (B5) at (4*.3,0); 
\diagram*{
   (I1) -- [ quarter right, looseness=.3,  thick] (J1)  -- [ quarter right, looseness=.3 ,  thick] (I1) ,
     (I1) -- [ half right, looseness=.8,  thick] (J1)  -- [ half right, looseness=.8 ,  thick] (I1) ,
   (I2) -- [ quarter right, looseness=.3,  thick] (J2)  -- [ quarter right, looseness=.3 ,  thick] (I2) ,
     (I2) -- [ half right, looseness=.8,  thick] (J2)  -- [ half right, looseness=.8 ,  thick] (I2) ,
   (I3) -- [ quarter right, looseness=.25,  thick] (J3)  -- [ quarter right, looseness=.25 ,  thick] (I3) ,
     (I3) -- [ half right, looseness=.55,  thick] (J3)  -- [ half right, looseness=.55 ,  thick] (I3) ,
        (A1) -- [scalar, thick, red] (B1) ,
        (A2) -- [scalar, thick, red] (B2) ,
       (A3) -- [scalar, thick, red] (B3) ,
       (A4) -- [scalar, thick, red] (B4) ,
       (A5) -- [scalar, thick, red] (B5) ,
};
  \end{feynman}
\end{tikzpicture}
  \end{minipage}  
  }

    \newcommand{\dnou}{
 \begin{minipage}[h]{0.1\linewidth}
\begin{tikzpicture}
\begin{feynman}[small]
 \node [dot] (I1) at (.45,1.49);
 \node [dot] (J1) at (.75,1.49);  
 \node [dot] (I2) at (.15,1.);
 \node [dot] (J2) at (1.05,1.);   
  \node [dot] (I3) at (-.15,.39);
 \node [dot] (J3) at (1.35,.39); 
   \vertex (A1) at (0,1.8); 
     \vertex (B1) at (0,0); 
   \vertex (A2) at (+.3,1.8); 
     \vertex (B2) at (+.3,0); 
   \vertex (A3) at (+2*.3,1.8); 
     \vertex (B3) at (+2*.3,0); 
   \vertex (A4) at (+3*.3,1.8); 
     \vertex (B4) at (3*.3,0); 
   \vertex (A5) at (4*.3,1.8); 
     \vertex (B5) at (4*.3,0); 
\diagram*{
   (I1) -- [ quarter right, looseness=.5,  thick] (J1)  -- [ quarter right, looseness=.5 ,  thick] (I1) ,
     (I1) -- [ half right, looseness=1.2,  thick] (J1)  -- [ half right, looseness=1.2 ,  thick] (I1) ,
   (I2) -- [ quarter right, looseness=.2,  thick] (J2)  -- [ quarter right, looseness=.2 ,  thick] (I2) ,
     (I2) -- [ half right, looseness=.65,  thick] (J2)  -- [ half right, looseness=.65 ,  thick] (I2) ,
   (I3) -- [ quarter right, looseness=.25,  thick] (J3)  -- [ quarter right, looseness=.25 ,  thick] (I3) ,
     (I3) -- [ half right, looseness=.55,  thick] (J3)  -- [ half right, looseness=.55 ,  thick] (I3) ,
        (A1) -- [scalar, thick, red] (B1) ,
        (A2) -- [scalar, thick, red] (B2) ,
       (A3) -- [scalar, thick, red] (B3) ,
       (A4) -- [scalar, thick, red] (B4) ,
       (A5) -- [scalar, thick, red] (B5) ,
};
  \end{feynman}
\end{tikzpicture}
  \end{minipage}  
  }

    \newcommand{\ctun}{
 \begin{minipage}[h]{0.08\linewidth}
\begin{tikzpicture}
\begin{feynman}[small]
 \node [dot] (I1) at (-.15,.8);
 \node [dot] (J1) at (.75,.8);  
 \node [dot] (I2) at (.45,0.3);
 \node [dot] (J2) at (1.05,0.3);  
   \vertex (A1) at (0,1.15); 
     \vertex (B1) at (0,0); 
   \vertex (A2) at (+.3,1.15); 
     \vertex (B2) at (+.3,0); 
   \vertex (A3) at (+2*.3,1.15); 
     \vertex (B3) at (+2*.3,0); 
   \vertex (A4) at (+3*.3,1.15); 
     \vertex (B4) at (3*.3,0); 
   \vertex (X1) at (0.15,1.1); 
     \vertex (X2) at (0.15,.5); 
\diagram*{  
   (I1) -- [ quarter right, looseness=.3,  thick] (J1)  -- [ quarter right, looseness=.3 ,  thick] (I1) ,
     (I1) -- [ half right, looseness=.8,  thick] (J1)  -- [ half right, looseness=.8 ,  thick] (I1) ,
   (I2) -- [ quarter right, looseness=.3,  thick] (J2)  -- [ quarter right, looseness=.3 ,  thick] (I2) ,
     (I2) -- [ half right, looseness=.8,  thick] (J2)  -- [ half right, looseness=.8 ,  thick] (I2) ,
        (A1) -- [scalar, thick, red] (B1) ,
        (A2) -- [scalar, thick, red] (B2) ,
       (A3) -- [scalar, thick, red] (B3) ,
       (A4) -- [scalar, thick, red] (B4) ,
       (X1) -- [ultra thick] (X2)  
};
  \end{feynman}
\end{tikzpicture}
  \end{minipage}  
  }

    \newcommand{\ctdos}{
 \begin{minipage}[h]{0.08\linewidth}
\begin{tikzpicture}
\begin{feynman}[small]
 \node [dot] (I1) at (.15,.8);
 \node [dot] (J1) at (1.05,.8);  
 \node [dot] (I2) at (-.15,0.3);
 \node [dot] (J2) at (.45,0.3);  
   \vertex (A1) at (0,1.15); 
     \vertex (B1) at (0,0); 
   \vertex (A2) at (+.3,1.15); 
     \vertex (B2) at (+.3,0); 
   \vertex (A3) at (+2*.3,1.15); 
     \vertex (B3) at (+2*.3,0); 
   \vertex (A4) at (+3*.3,1.15); 
     \vertex (B4) at (3*.3,0); 
   \vertex (X1) at (0.75,1.1); 
     \vertex (X2) at (0.75,.5); 
\diagram*{  
   (I1) -- [ quarter right, looseness=.3,  thick] (J1)  -- [ quarter right, looseness=.3 ,  thick] (I1) ,
     (I1) -- [ half right, looseness=.8,  thick] (J1)  -- [ half right, looseness=.8 ,  thick] (I1) ,
   (I2) -- [ quarter right, looseness=.3,  thick] (J2)  -- [ quarter right, looseness=.3 ,  thick] (I2) ,
     (I2) -- [ half right, looseness=.8,  thick] (J2)  -- [ half right, looseness=.8 ,  thick] (I2) ,
        (A1) -- [scalar, thick, red] (B1) ,
        (A2) -- [scalar, thick, red] (B2) ,
       (A3) -- [scalar, thick, red] (B3) ,
       (A4) -- [scalar, thick, red] (B4) ,
       (X1) -- [ultra thick] (X2)  
};
  \end{feynman}
\end{tikzpicture}
  \end{minipage}  
  }

    \newcommand{\cttres}{
 \begin{minipage}[h]{0.08\linewidth}
\begin{tikzpicture}
\begin{feynman}[small]
 \node [dot] (I1) at (-.15,.8);
 \node [dot] (J1) at (1.05,.8);  
 \node [dot] (I2) at (.15,0.3);
 \node [dot] (J2) at (.45,0.3);  
   \vertex (A1) at (0,1.15); 
     \vertex (B1) at (0,0); 
   \vertex (A2) at (+.3,1.15); 
     \vertex (B2) at (+.3,0); 
   \vertex (A3) at (+2*.3,1.15); 
     \vertex (B3) at (+2*.3,0); 
   \vertex (A4) at (+3*.3,1.15); 
     \vertex (B4) at (3*.3,0); 
   \vertex (X1) at (0.75,1.1); 
     \vertex (X2) at (0.75,.5); 
\diagram*{  
   (I1) -- [ quarter right, looseness=.2,  thick] (J1)  -- [ quarter right, looseness=.2 ,  thick] (I1) ,
     (I1) -- [ half right, looseness=.45,  thick] (J1)  -- [ half right, looseness=.45 ,  thick] (I1) ,
   (I2) -- [ quarter right, looseness=.5,  thick] (J2)  -- [ quarter right, looseness=.5 ,  thick] (I2) ,
     (I2) -- [ half right, looseness=1.2,  thick] (J2)  -- [ half right, looseness=1.2 ,  thick] (I2) ,
        (A1) -- [scalar, thick, red] (B1) ,
        (A2) -- [scalar, thick, red] (B2) ,
       (A3) -- [scalar, thick, red] (B3) ,
       (A4) -- [scalar, thick, red] (B4) ,
       (X1) -- [ultra thick] (X2)  
};
  \end{feynman}
\end{tikzpicture}
  \end{minipage}  
  }

    \newcommand{\ctquatre}{
 \begin{minipage}[h]{0.08\linewidth}
\begin{tikzpicture}
\begin{feynman}[small]
 \node [dot] (I1) at (-.15,.8);
 \node [dot] (J1) at (1.05,.8);  
 \node [dot] (I2) at (.45,0.3);
 \node [dot] (J2) at (.75,0.3);  
   \vertex (A1) at (0,1.15); 
     \vertex (B1) at (0,0); 
   \vertex (A2) at (+.3,1.15); 
     \vertex (B2) at (+.3,0); 
   \vertex (A3) at (+2*.3,1.15); 
     \vertex (B3) at (+2*.3,0); 
   \vertex (A4) at (+3*.3,1.15); 
     \vertex (B4) at (3*.3,0); 
   \vertex (X1) at (0.15,1.1); 
     \vertex (X2) at (0.15,.5); 
\diagram*{  
   (I1) -- [ quarter right, looseness=.2,  thick] (J1)  -- [ quarter right, looseness=.2 ,  thick] (I1) ,
     (I1) -- [ half right, looseness=.45,  thick] (J1)  -- [ half right, looseness=.45 ,  thick] (I1) ,
   (I2) -- [ quarter right, looseness=.5,  thick] (J2)  -- [ quarter right, looseness=.5 ,  thick] (I2) ,
     (I2) -- [ half right, looseness=1.2,  thick] (J2)  -- [ half right, looseness=1.2 ,  thick] (I2) ,
        (A1) -- [scalar, thick, red] (B1) ,
        (A2) -- [scalar, thick, red] (B2) ,
       (A3) -- [scalar, thick, red] (B3) ,
       (A4) -- [scalar, thick, red] (B4) ,
       (X1) -- [ultra thick] (X2)  
};
  \end{feynman}
\end{tikzpicture}
  \end{minipage}  
  }

    \newcommand{\ctcinc}{
 \begin{minipage}[h]{0.06\linewidth}
\begin{tikzpicture}
\begin{feynman}[small]
 \node [dot] (I1) at (-.15,.3);
 \node [dot] (J1) at (.75,.3);  
   \vertex (A1) at (0,0.65); 
     \vertex (B1) at (0,-.05); 
   \vertex (A2) at (+.3,0.65); 
     \vertex (B2) at (+.3,-.05); 
   \vertex (A3) at (+2*.3,0.65); 
     \vertex (B3) at (+2*.3,-0.05);  
   \vertex (X1) at (0.15,.61); 
     \vertex (X2) at (0.15,.005); 
   \vertex (X3) at (0.45,.61); 
     \vertex (X4) at (0.45,.005); 
\diagram*{  
   (I1) -- [ quarter right, looseness=.35,  thick] (J1)  -- [ quarter right, looseness=.35 ,  thick] (I1) ,
     (I1) -- [ half right, looseness=.7,  thick] (J1)  -- [ half right, looseness=.7 ,  thick] (I1) ,
        (A1) -- [scalar, thick, red] (B1) ,
        (A2) -- [scalar, thick, red] (B2) ,
       (A3) -- [scalar, thick, red] (B3) ,
       (X1) -- [ultra thick] (X2)  ,
       (X3) -- [ultra thick] (X4)  ,
};
  \end{feynman}
\end{tikzpicture}
  \end{minipage}  
  }

    \newcommand{\ctsis}{
 \begin{minipage}[h]{0.08\linewidth}
\begin{tikzpicture}
\begin{feynman}[small]
 \node [dot] (I1) at (-.15,.8);
 \node [dot] (J1) at (1.05,.8);  
 \node [dot] (I2) at (.15,0.3);
 \node [dot] (J2) at (.75,0.3);  
   \vertex (A1) at (0,1.15); 
     \vertex (B1) at (0,0); 
   \vertex (A2) at (+.3,1.15); 
     \vertex (B2) at (+.3,0); 
   \vertex (A3) at (+2*.3,1.15); 
     \vertex (B3) at (+2*.3,0); 
   \vertex (A4) at (+3*.3,1.15); 
     \vertex (B4) at (3*.3,0); 
   \vertex (X1) at (0.45,1.1); 
     \vertex (X2) at (0.45,.01); 
\diagram*{  
   (I1) -- [ quarter right, looseness=.2,  thick] (J1)  -- [ quarter right, looseness=.2 ,  thick] (I1) ,
     (I1) -- [ half right, looseness=.45,  thick] (J1)  -- [ half right, looseness=.45 ,  thick] (I1) ,
   (I2) -- [ quarter right, looseness=.3,  thick] (J2)  -- [ quarter right, looseness=.3 ,  thick] (I2) ,
     (I2) -- [ half right, looseness=.8,  thick] (J2)  -- [ half right, looseness=.8 ,  thick] (I2) ,
        (A1) -- [scalar, thick, red] (B1) ,
        (A2) -- [scalar, thick, red] (B2) ,
       (A3) -- [scalar, thick, red] (B3) ,
       (A4) -- [scalar, thick, red] (B4) ,
       (X1) -- [ultra thick] (X2)  
};
  \end{feynman}
\end{tikzpicture}
  \end{minipage}  
  }
  
    \newcommand{\ctset}{
 \begin{minipage}[h]{0.08\linewidth}
\begin{tikzpicture}
\begin{feynman}[small]
 \node [dot] (I1) at (.15,1);
 \node [dot] (J1) at (1.05,1);  
 \node [dot] (I2) at (-.15,0.35);
 \node [dot] (J2) at (.75,0.35);  
   \vertex (A1) at (0,1.33); 
     \vertex (B1) at (0,0); 
   \vertex (A2) at (+.3,1.33); 
     \vertex (B2) at (+.3,0); 
   \vertex (A3) at (+2*.3,1.33); 
     \vertex (B3) at (+2*.3,0); 
   \vertex (A4) at (+3*.3,1.33); 
     \vertex (B4) at (3*.3,0); 
   \vertex (X1) at (0.45,1.31); 
     \vertex (X2) at (0.45,.02); 
\diagram*{  
   (I1) -- [ quarter right, looseness=.3,  thick] (J1)  -- [ quarter right, looseness=.3 ,  thick] (I1) ,
     (I1) -- [ half right, looseness=.8,  thick] (J1)  -- [ half right, looseness=.8 ,  thick] (I1) ,
   (I2) -- [ quarter right, looseness=.3,  thick] (J2)  -- [ quarter right, looseness=.3 ,  thick] (I2) ,
     (I2) -- [ half right, looseness=.8,  thick] (J2)  -- [ half right, looseness=.8 ,  thick] (I2) ,
        (A1) -- [scalar, thick, red] (B1) ,
        (A2) -- [scalar, thick, red] (B2) ,
       (A3) -- [scalar, thick, red] (B3) ,
       (A4) -- [scalar, thick, red] (B4) ,
       (X1) -- [ultra thick] (X2)  
};
  \end{feynman}
\end{tikzpicture}
  \end{minipage}  
  }

    \newcommand{\ctvdos}{
 \begin{minipage}[h]{0.08\linewidth}
\begin{tikzpicture}
\begin{feynman}[small]
 \node [dot] (I1) at (0,.35);
 \node [dot] (J1) at (1,.35);  
   \vertex (A1) at (.25,.8); 
     \vertex (B1) at (.25,-.15); 
     \vertex (B1t) at (.25,-.32)  {{{\roig \scriptsize{$^{E_1}$}}}}; 
   \vertex (A2) at (+.75,.8); 
     \vertex (B2) at (+.75,-.15); 
        \vertex (B2t) at (.75,-.32)  {{{\roig \scriptsize{$^{E_1}$}}}}; 
   \vertex (X1) at (0.5,.7); 
     \vertex (X2) at (0.5,.0); 
      \vertex (X1t) at (1.,1)  { \scriptsize{$^{K_2}$}}; 
       \vertex (Y1) at (.75,1.05); 
     \vertex (Y2) at (0.5,.8); 
\diagram*{  
   (I1) -- [ quarter right, looseness=.3,  thick] (J1)  -- [ quarter right, looseness=.3 ,  thick] (I1) ,
     (I1) -- [ half right, looseness=.8,  thick] (J1)  -- [ half right, looseness=.8 ,  thick] (I1) ,
        (A1) -- [scalar, thick, red] (B1) ,
        (A2) -- [scalar, thick, red] (B2) ,
       (X1) -- [ultra thick] (X2) ,
        (Y1) -- [   quarter right, -stealth] (Y2)  ,
};
  \end{feynman}
\end{tikzpicture}
  \end{minipage}  
  }

\title{Effective Hamiltonians and Counterterms \\[.2cm] for Hamiltonian  Truncation}

\author{Joan Elias Mir\'o, }

\affiliation{Abdus Salam International Centre for Theoretical Physics, Strada Costiera 11, 34151, Trieste, Italy}

\author{James Ingoldby}

\date{\today}

\abstract{We outline a procedure for applying Hamiltonian Truncation to Quantum Field Theories (QFTs) that have UV divergences. To do this, we  derive a novel representation of an Effective Hamiltonian which makes manifest some of its important properties (e.g.  the non-perturbative matching of the spectra between the UV theory and the theory described by the Effective Hamiltonian). We check the consistency of   our procedure  using  Conformal Perturbation Theory. Finally we comment on how the Effective Hamiltonian, which incorporates  non-local interactions,  describes a local QFT.}

\begin{document}

\maketitle
\flushbottom

\setcounter{tocdepth}{1}

\section{Introduction}
\label{sec:innn}

Increasing the variety and scope of theoretical tools that can be used to analyse strongly coupled Quantum Field Theories (QFTs) reliably remains an important scientific challenge. Hamiltonian Truncation is one such tool. To apply it, the Hamiltonian of the full QFT (which is infinite-dimensional) is approximated by a finite dimensional Hamiltonian which acts only on the lowest energy states of the theory. This finite dimensional Hamiltonian can then be diagonalised numerically, enabling the low-energy spectrum to be estimated.
	
Hamiltonian Truncation was first applied in the context of QFT in Ref.~\cite{Brooks:1984aaa}, and subsequently it was successfully employed in Refs.~\cite{Yurov:1989yu,Yurov:1991my} to non-perturbatively determine renormalisation group flows between CFTs with $d=1+1$ on the cylinder. Since then, the method has been generalised to QFTs with arbitrary $d$ in \cite{Hogervorst:2014rta} and applied to QFTs quantised on the lightcone, beginning with \cite{Brodsky:1985aaa}. In recent years, there has been a resurgence of interest in Hamiltonian Truncation within high energy physics. See for example Refs.~\cite{Bajnok:2015bgw,Katz:2016hxp,Anand:2017yij,Rutter:2018aog,Fitzpatrick:2018ttk,Hogervorst:2018otc,Delacretaz:2018xbn,Anand:2019lkt,Fitzpatrick:2019cif,Elias-Miro:2020qwz,Anand:2020gnn,Anand:2020qnp,Hogervorst:2021spa,Anand:2021qnd,EliasMiro:2021aof,Szasz-Schagrin:2022wkk,Chen:2022zms,Delacretaz:2022ojg,Henning:2022xlj,Dempsey:2022uie}, as well as the reviews \cite{Fitzpatrick:2022dwq,Konik-review}. A particularly important focus for research in this area has been on the development of effective Hamiltonians for improving the accuracy of Hamiltonian Truncation estimates~\cite{Hogervorst:2014rta,Giokas:2011ix,Rychkov:2014eea,Rychkov:2015vap,Elias-Miro:2015bqk,Elias-Miro:2017xxf,Elias-Miro:2017tup,Cohen:2021erm}.
	

However, the absence of a systematic understanding of how Hamiltonian Truncation should be applied to QFTs with UV divergences (which require renormalisation) has been a significant obstacle, preventing wider usage of the method. When the Hamiltonian is truncated, removing high energy states from the QFT, this acts as a UV regulator. In Ref.~\cite{EliasMiro:2021aof}, we showed that the Hamiltonian Truncation UV regulator gives results that are inconsistent with those derived using a local regularisation (such as a short distance cutoff), implying that non-local counterterms are needed to implement renormalisation in Hamiltonian Truncation.

In this work, we present three main results: First in Section~\ref{sec:heff} we derive a novel representation of an effective Hamiltonian that is finite dimensional, but constructed to have the exact same spectrum of lowest energy states as the full QFT. We then compare it with alternative types of effective Hamiltonian from the literature.

Second, we show in Section~\ref{sec:hampt} how an effective Hamiltonian can be used to compute the spectrum of QFTs which have UV divergences. This is done by employing a local regularisation and introducing the corresponding local counterterms to remove all UV divergences before building an effective Hamiltonian for the renormalised QFT, which can be diagonalised numerically.
We provide   explicit formulae for this Effective Hamiltonian   in Section~\ref{tcsaex}.

Our third result is presented in Section~\ref{sec:cutoffdep}. Here we perform a check using perturbation theory that the effective Hamiltonian constructed following the procedure outlined in Section~\ref{sec:hampt} contains the extra interactions required to cancel the UV divergences found in Ref.~\cite{EliasMiro:2021aof} which were inconsistent with local regularisation.

\section{Effective Hamiltonians}
\label{sec:heff}

In this section, we first summarise how effective Hamiltonians can be used to improve the accuracy of Hamiltonian Truncation calculations~\cite{Hogervorst:2014rta,Giokas:2011ix,Rychkov:2014eea,Rychkov:2015vap,Elias-Miro:2015bqk,Elias-Miro:2017xxf,Elias-Miro:2017tup,Cohen:2021erm}. We then present a novel representation for a particular effective Hamiltonian, which makes its useful properties manifest.

The start point for any Hamiltonian Truncation calculation is reached by first expressing the Hamiltonian in a well chosen basis and dividing  it into a solvable part $H_0$ and a deformation $V$,  
\be
H=H_0+V \, .   \label{ham1}
\ee 
The solvable part has the spectrum $H_0|i\rangle = E_i |i\rangle$.
We are interested in finding the low energy spectrum
of the interacting theory,
\be
H|\Eps\rangle = \Eps_l |\Eps \rangle  \, . \label{eigqft}
\ee
where $ \Eps_l $ is the  interacting energy of a low energy state. We emphasise that the Hilbert space the Hamiltonian above acts upon is infinite dimensional.

Next, we proceed  by separating  the Hilbert space  into a low  ($l$) and high ($h$)  energy sector
${\cal H}={\cal H}_l\oplus {\cal H}_h$. The low energy Hilbert space 
${\cal H}_l$ is finite dimensional and  spanned by the states $\sum_{E_i \leq E_T}|i \rangle \langle i | $; while the high energy Hilbert space ${\cal H}_h$ is 
infinite dimensional and contains the rest of the states $\sum_{E_i > E_T}|i \rangle \langle i | $.
The constant $E_T$ is called the Hamiltonian Truncation cutoff. 

The calculation proceeds by replacing the infinite dimensional eigenvalue problem in \reef{eigqft} with an equivalent finite dimensional eigenvalue problem
\be
H_\text{eff}|\Eps_l\rangle = \Eps_l |\Eps_l \rangle  \, , \label{heftone}
\ee
where $|\Eps_l \rangle\in {\cal H}_l$. 
The Effective Hamiltonian operator $H_\text{eff}$ acts in the low energy Hilbert space only. If the truncation cutoff is much larger than the energy of the interacting states that we are interested in so that $E_T \gg \Eps_l$, then 
\be
\langle i|H_\text{eff}|j\rangle \approx \langle i | H | j \rangle  \, ,  \label{rawapp}
\ee 
where $| i \rangle\, , | j \rangle$  are $H_0$ eigenstates in the subspace $\calH_l$. This approximation is often called \emph{raw} Hamiltonian Truncation. So long as the eigenvalues of $H$ are all finite, in the limit of large Hamiltonian Truncation cutoff, we would recover the spectrum of the interacting theory exactly. Even for QFT Hamiltonians which are unbounded from above, the approximation in \reef{rawapp} can also determine the lowest energy eigenvalues, if the QFT has no UV divergences.

However it is often hard to use sufficiently large values for the cutoff $E_T$ in practical calculations, because the dimension of the low energy Hilbert space ${\cal H}_l$  grows exponentially with $E_T$. It is thus desirable to find better approximations to $H_\text{eff}$  than  \reef{rawapp}, without increasing $E_T$.

In perturbation theory, the effective Hamiltonian is expressed as a series
\begin{align}
	H_\text{eff} = H_0 + V + \sum_{n=2}^\infty H_{\text{eff}\, n}\,.\label{eq:heffgen}
\end{align}
where $H_{\text{eff}\, n}=O(V^n)$.~\footnote{We remark that the approximation introduced by truncating this series to fixed order in $V$ breaks down at strong coupling, but that it is possible to find alternative rigorous approximations to $H_\text{eff}$ that are valid at strong coupling and agree with \reef{eq:heffgen} at weak coupling~\cite{Elias-Miro:2017xxf,Elias-Miro:2017tup}.}
The effective Hamiltonian is not unique, and different effective Hamiltonians are related to one another by similarity transformations.
Therefore several methods are possible to determine \reef{eq:heffgen}.
In the next section we will employ the effective Hamiltonian $H_\text{eff}$ introduced in Ref.~\cite{Cohen:2021erm}, which we shall refer to as the CFHL effective Hamiltonian. In appendix \ref{appalter} we comment on  alternative $H_\text{eff}$ constructions. 
The first few terms of the series \reef{eq:heffgen} for the CFHL effective Hamiltonian are given by
\begin{align}
\left(H_{\text{eff}\, 2}\right)_{fi} = \frac{V_{fh}V_{hi}}{E_{fh}}\,,
\label{heff2}	
\end{align}
where we have introduced the notation $X_{ab}\equiv\matrixel{a}{X}{b}$ for any operator $X$, and $E_{ab}\equiv E_a-E_b$. The states $\ket{f}$, $\ket{i}$ are $H_0$ eigenstates in the subspace $\calH_l$, and the repeated index $h$ indicates summation over all $H_0$ eigenstates with energies \emph{above} the HT cutoff $E_T$. 
The third order expression was found to be
\begin{align}
	\left(H_{\text{eff}\,3}\right)_{fi}=\frac{V_{fh_1}V_{h_1h_2}V_{h_2i}}{E_{fh_1}E_{fh_2}}-\frac{V_{fl}V_{lh}V_{hi}}{E_{fh}E_{lh}}\,.\label{heff3}
\end{align}
For future reference, we  also provide  the CFHL $H_\text{eff}$ at fourth order in perturbation theory, 
\begin{align}
	\left(H_{\text{eff}\, 4}\right)_{fi}=&\frac{V_{fh_1}V_{h_1h_2}V_{h_2h_3}V_{h_3i}}{E_{fh_1}E_{fh_2}E_{fh_3}}-\frac{V_{fh_1}V_{h_1l}V_{lh_2}V_{h_2i}}{E_{fh_1}E_{fh_2}E_{lh_2}} \nonumber\\
	&-\frac{V_{fl}V_{lh_1}V_{h_1h_2}V_{h_2i}}{E_{fh_1}E_{lh_2}}\left[\frac{1}{E_{fh_2}}+\frac{1}{E_{lh_1}}\right]+\frac{V_{fl_1}V_{l_1l_2}V_{l_2h}V_{hi}}{E_{fh}E_{l_1h}E_{l_2h}}\, .
	\label{heff4}
\end{align}


\subsection{Alternative Representation of the Effective  Hamiltonian}

We now present an alternative formula for the CFHL effective Hamiltonian~\cite{Cohen:2021erm} in terms of the diagonal Hamiltonian $H_0$ and deformation $V$. We then show that the effective Hamiltonian has the same spectrum as the full Hamiltonian for states with energies beneath the HT cutoff.

Our alternative formula depends on the time evolution operator in the interaction picture. For the theory with Hamiltonian given by \reef{ham1}, this operator is
	\begin{align}
		U_\text{IP}(t_f,t_i) = T\exp\left\{-i\int_{t_i}^{t_f}dt \, V_\text{IP}(t)\right\}\, , \qquad V_\text{IP}(t) = e^{i H_0 t}V e^{-\eta t} e^{-i H_0 t}\, .
		\label{eq:timeevol}
	\end{align}
	Here $\eta$ is a parameter that controls the adiabatic switching off of $V$ at large positive times. The symbol $\Sigma$ denotes the large time limit of this operator
	\begin{align}
		\Sigma = \lim\limits_{t_f\rightarrow\infty} U_\text{IP}(t_f,0)\, .
		\label{eq:sigdef}
	\end{align}
We employ the notation $X_l$ to refer to the truncation of an operator $X$ to include only its matrix elements between states in the subspace ${\cal H}_l$. Finally, our alternative formula for the effective Hamiltonian is
\be
	 H_\text{eff}=\left(\Sigma_{l}\right)^{-1}\left[\Sigma(H_0+V)\Sigma^{\dagger}\right]_l\Sigma_{l}\,,\label{hl2neat}
\ee
which is the main result in this section. The equation above fully determines the effective Hamiltonian in terms of $H_0$ and $V$. 	

\begin{figure}
	\centering
	\includegraphics[width=0.5\columnwidth]{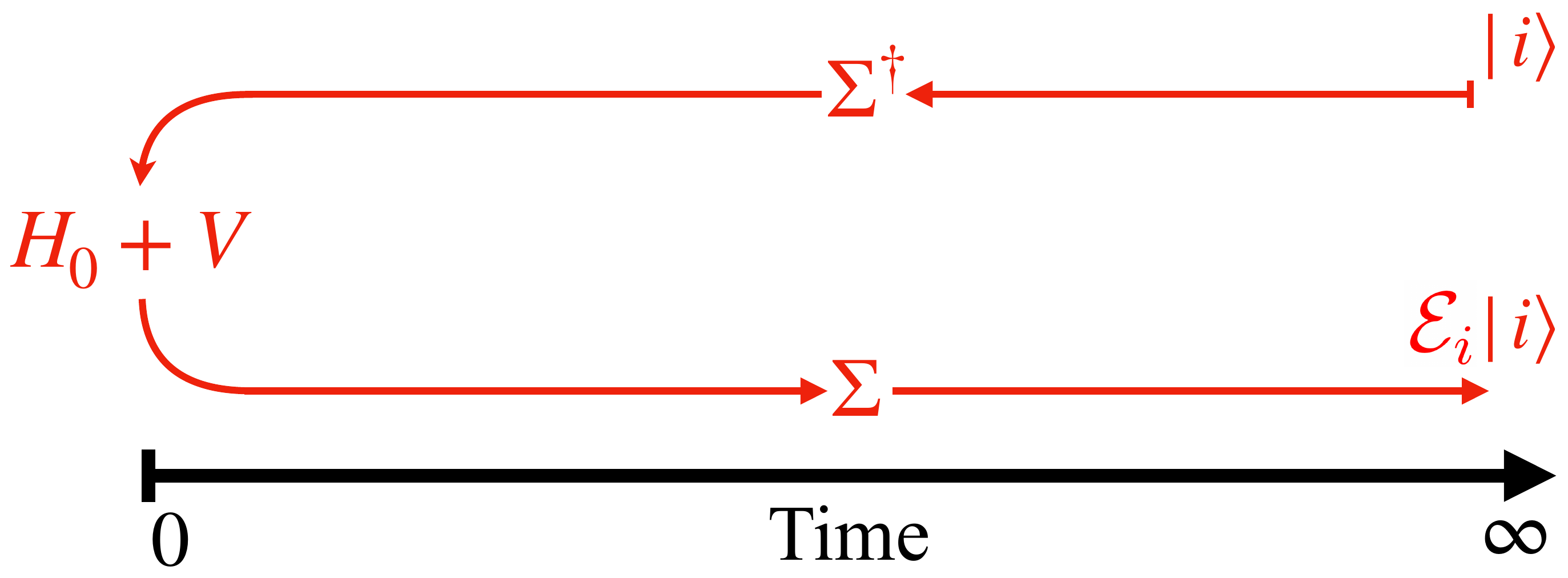}
	\caption{Cartoon describing the action of the product of operators $\Sigma\left(H_0+V\right)\Sigma^\dagger$ on an eigenstate of $H_0$ denoted $\ket{i}$. First, $\Sigma^\dagger$ evolves $\ket{i}$ back from future infinity to an eigenstate of $H_0+V$ at zero time. Then, by acting on that eigenstate with $H_0+V$ we bring down a factor of $\calE_i$. Finally $\Sigma$ evolves the state forward in time, undoing the action of $\Sigma^\dagger$.}
	\label{timeevo}
\end{figure}

The product of operators $\Sigma\left(H_0+V\right)\Sigma^\dagger$ in \reef{hl2neat} acts particularly simply on eigenstates of $H_0$ -- see Fig.~\ref{timeevo} for a pictorial representation. The time evolution operator $\Sigma^\dagger$ takes these eigenstates at time $t_f\rightarrow\infty$ (which are also eigenstates of the full Hamiltonian due to the $e^{-\eta t}$ factor shown in \reef{eq:timeevol} that switches off the deformation at large times) and evolves them back to states at zero time. During this evolution, the full Hamiltonian of the theory will change from $H_0$ to $H_0+V$. This evolution will be adiabatic if the parameter $\eta$ is sufficiently small.~\footnote{For simplicity we assume  that $H_0$ has no degeneracies so that the adiabatic theorem can be straightforwardly applied, but it is possible to relax this assumption and generalise our derivation.} Using the adiabatic theorem, we conclude that $\Sigma^\dagger$ evolves eigenstates of the full Hamiltonian at infinite time $H_0$ to eigenstates of the full Hamiltonian at zero time $H_0+V$.

If we act with operator $H_0+V$ at $t=0$, the energy eigenstate will be multiplied by $\calE_i$, its corresponding energy eigenvalue. Finally if we act with $\Sigma$, we then adiabatically time evolve back to infinite future time to yield a state proportional to the $H_0$ eigenstate we started with. In summary, we expect
\begin{align}
	\lim\limits_{\eta\rightarrow0}\Sigma\left(H_0+V\right)\Sigma^\dagger\sim\text{diag}\left(\calE_0,\calE_1,\dots\right)\,,\label{eq:sigintuition}
\end{align}
when the operator on the left hand side is expressed in the eigenbasis of $H_0$. Unfortunately, this expectation is not always met because $\Sigma$ can become singular in the $\eta\rightarrow 0$ limit. Nonetheless we conclude that $\Sigma$ plays a special role: when it is free of divergences, it diagonalises the Hamiltonian $H_0+V$.
Once the full Hamiltonian is diagonalised as  in \reef{eq:sigintuition}, its matrix elements between states outside the subspace $\calH_l$ can be thrown away without losing information about the low energy eigenstates and eigenvalues of the full Hamiltonian. 

To obtain a heuristic derivation of \reef{hl2neat}, 
we start from an effective Hamiltonian $H_\text{eff}=H_0+V_\text{eff}$ acting on $\calH_l$, and out of $H_\text{eff}$ we construct an analogous time evolution operator $\Sigma_\text{eff}$:
	\begin{align}
		\Sigma_\text{eff} = \lim\limits_{t_f\rightarrow\infty}T\exp\left\{-i\int_{0}^{t_f}dt\, V_{\text{eff IP}}(t)\right\}\, , \quad V_{\text{eff IP}}(t) = e^{i H_0 t}V_\text{eff} e^{-\eta t} e^{-i H_0 t}\, .
		\label{eq:sigeffdef}
	\end{align}
This Effective Hamiltonian is also diagonalised by its time evolution operator $\Sigma_{\text{eff}}$. Next, following CFHL~\cite{Cohen:2021erm}, we demand that the matrix elements of the time evolution operators between the full \reef{ham1} and the effective field theory should be matched
	\begin{align}
		\Sigma_{\text{eff}} = \Sigma_l\, .
	\end{align}
Assuming that the spectra of the full and effective Hamiltonians are also matched by matching time evolution operators, we are
led to the following equation for an effective Hamiltonian 
\begin{align}
	\Sigma_l H_\text{eff}(\Sigma_l)^{-1} = \left[\Sigma(H_0+V)\Sigma^\dagger\right]_l \, ,
	\label{eq:hl2}
\end{align}
which is equivalent to \reef{hl2neat}. Since $\Sigma_l$ is non--unitary as a result of its projection, we note that $H_\text{eff}$ is non--hermitian.

We now comment further on the origin of singularities in $\Sigma$ and provide a more rigorous proof that the CFHL Hamiltonian is given by \reef{hl2neat}. Physically, these singularities arise if eigenstates of $H_0+V$ pick up large phases $\exp\{i(\Delta E)/\eta\}$ in the $\eta\rightarrow0$ limit under time evolution by $\Sigma$. Interaction picture energy eigenstates acquire phases during time evolution when there is a difference between the energy eigenvalues of the free and interacting theories, so that $\Delta E\sim \calE_i-E_i$.

In scattering problems, at infinite volume, the time evolution operator $\Sigma$ is commonly referred to as a M{\o}ller operator \cite{kleinert2016particles}.
In this context, energy eigenstates of the interacting theory are states of many particles which are distantly separated from one another at time $t_f\rightarrow\infty$ and have vanishing mutual interactions in this limit. As a result, the energy eigenvalues of the interacting theory match those of a theory of 
free particles and $\Delta E=0$. The scattering states of the interacting theory may be constructed from free particle states using the Lippmann-Schwinger equation, thus yielding a well defined $\Sigma$.~\footnote{The M{\o}ller operator in \reef{eq:timeevol} determines the S-matrix $S=U_\text{IP}(+\infty,0) U_\text{IP}^\dagger(-\infty,0)$. 
This operator may suffer from IR divergences in gauge theories however.}

In the context of a general quantum theory in finite volume, the $H_0+V$ eigenvalues will not match those of $H_0$. However, in this case eigenstates of $H_0+V$ may still be constructed from time evolution operators without introducing a singular phase using the first Gell--Mann-Low formula \cite{Gell-Mann:1951ooy}, which is
\begin{align}
	\ket{\psi_i} = \lim_{\eta\rightarrow 0} \frac{U_{IP}^{-1}(\infty,0)\ket{i}}{\bra{i}{U_{IP}^{-1}(\infty,0)}\ket{i}}\,,\label{eq:gm1a}
\end{align}
where $\ket{\psi_i}$ is an eigenstate of $H_0+V$ and $\ket{i}$ is an $H_0$ eigenstate. The time evolution operators $U_{IP}$ depend on the switching parameter $\eta$. Both the numerator and denominator are proportional to the phase factor $\exp\{i(\Delta E)/\eta\}$, but the ratio has a well defined limit. Similarly, we have
\begin{align}
	\bra{\psi_i} = \lim_{\eta\rightarrow 0}\frac{\bra{i}U_{IP}(\infty,0)}{\bra{i}U_{IP}(\infty,0)\ket{i}}\,.
	\label{eq:gm1b}
\end{align}

By replacing $U_{IP}$ with $\Sigma$ in expressions \reef{eq:gm1a} and \reef{eq:gm1b}, the interacting theory Hamiltonian may be diagonalised to yield
\begin{align}
	\lim_{\eta\rightarrow 0}\frac{\matrixel{f}{\Sigma(H_0+V)\Sigma^\dagger}{i}}{\bra{f}\Sigma\ket{f}\bra{i}\Sigma^\dagger\ket{i}} = \delta_{fi}\lim_{\eta\rightarrow 0}\frac{{\mathcal E}_i}{\bra{i}\Sigma\ket{i}\bra{i}{\Sigma^\dagger}\ket{i}}\, ,
\end{align}
where ${\mathcal E}_i$ are the energy eigenvalues of the interacting theory. The individual factors in the denominators all contain phases that diverge in the $\eta\rightarrow0$ limit, but the ratios on both sides of the equation have a well defined limit. 

If we restrict attention to states $\ket{i},\ket{f} \in {\cal H}_l$, i.e. in the low energy Hilbert space, we find
\begin{align}
	\lim_{\eta\rightarrow 0}\frac{\matrixel{f}{\left[\Sigma(H_0+V)\Sigma^\dagger\right]_l}{i}}{\bra{f}\Sigma_l\ket{f}\bra{i}\Sigma_l^\dagger\ket{i}} = \delta_{fi}\lim_{\eta\rightarrow 0}\frac{{\mathcal E}_{l,i}}{\matrixel{i}{\Sigma_l}{i}\bra{i}\Sigma_l^\dagger\ket{i}}\,,
	\label{eq:fulldiag}
\end{align}
where ${\mathcal E}_{l,i}$ refer to the energy eigenvalues of states connected to ${\cal H}_l$ through \reef{eq:gm1a}. The ${\mathcal E}_{l,i}$ are also given by the Rayleigh--Schr{\"o}dinger (RS) perturbative series. Therefore, we could generate the RS series by expanding \reef{eq:fulldiag} perturbatively. We comment further on the connection with RS perturbation theory in Appendix~\ref{app:rs}.

The Hamiltonian of the low energy effective theory may also be diagonalised using Eqs.~\reef{eq:gm1a} and \reef{eq:gm1b}
\begin{align}
	\lim_{\eta\rightarrow 0}\frac{\matrixel{f}{\Sigma_\text{eff}H_\text{eff}\Sigma_\text{eff}^{-1}}{i}}{\matrixel{f}{\Sigma_\text{eff}}{f}\bra{i}\Sigma^{-1}_\text{eff}\ket{i}} = \delta_{fi}\lim_{\eta\rightarrow 0}\frac{{\mathcal E}_{\text{eff},i}}{\matrixel{i}{\Sigma_\text{eff}}{i}\bra{i}\Sigma^{-1}_\text{eff}\ket{i}}\, .
	\label{eq:effdiag}
\end{align}
The ${\mathcal E}_{\text{eff},i}$ refer to the energy eigenvalues of the effective theory. Note also that $\Sigma_{\text{eff}}^{-1}\neq\Sigma_{\text{eff}}^\dagger$ in the effective theory. For this reason, we replace $U_{IP}^{-1}(\infty,0)$ with $\Sigma_{\text{eff}}^{-1}$ when using \reef{eq:gm1a}.

Energy eigenvalues are also determined by time evolution operators, through the second Gell--Mann-Low formula \cite{Gell-Mann:1951ooy}, which may be rewritten using our notation as:
\begin{align}
	{\mathcal E}_i - E_i = \lim_{\eta\rightarrow0} \, -i\eta g \pd{\,}{g}\log\left\{\bra{i}{\Sigma}\ket{i}\right\}\, ,
	\label{eq:gm2}
\end{align}
where $g$ is a coupling strength proportional to the interaction $V\propto g$, and ${\mathcal E}_i - E_i$ represents the shift in energy between the full and solvable $H_0$ theory.

In the case of the CFHL effective Hamiltonian, matrix elements of the time evolution operator are matched $\Sigma_l=\Sigma_{\text{eff}}$. Since the energy eigenvalues are totally determined by the time evolution operator through \reef{eq:gm2}, it follows that the energy eigenvalues of the CFHL Hamiltonian will match exactly those of the full theory which are connected to states in ${\cal H}_l$, so that ${\mathcal E}_{\text{eff},i}={\mathcal E}_{l,i}$.

We may therefore combine Eqs.~(\ref{eq:fulldiag}) and (\ref{eq:effdiag}) to obtain \reef{eq:hl2}, after canceling common factors of $\Sigma_l=\Sigma_{\text{eff}}$ and taking the $\eta\rightarrow0$ limit to get finite expressions on both sides. The result can be rearranged to yield \reef{hl2neat}.
We emphasise that the energy eigenvalues of $H_\text{eff}$ are guaranteed to match (a subset of) those of $H_0+V$. This follows from the matching condition $\Sigma_{\text{eff}}=\Sigma_l$ and the second Gell--Mann-Low equation \reef{eq:gm2}.
Expanding \reef{hl2neat} perturbatively in $V$  yields (\ref{heff2})--(\ref{heff4}).

Further difficulties may be encountered when computing matrix elements of a Hermitian observable $A$, other than the Hamiltonian $H+V$, in the low-energy Hilbert space. The Effective  Hamiltonian in  \reef{hl2neat} is not Hermitian $H_\text{eff}^\dagger\neq H_\text{eff}$, thus its eigenvectors are not expected to be orthogonal. Therefore these states are not convenient to compute the matrix elements of the operator $A$ in the low energy effective Hilbert space. 
To compute matrix elements we should use the time-evolution operator,
\be
A_\text{eff}=\left(\Sigma_{l}\right)^{-1}\left[\Sigma   A \Sigma^{\dagger}\right]_l\Sigma_{l}\, ,
\ee
with matrix elements defined only in  the low energy Hilbert space ${\cal H}_l$. The first few orders are given by
$ ( A_\text{eff})_{fi}=A_{fi}+ \frac{V_{fh} A_{hi}}{E_{fh}}+ \frac{A_{fh} V_{hi}}{E_{ih}}+ O(V^2)$.

We observe that \reef{hl2neat} is constructed as a similarity transformation of the full Hamiltonian which diagonalises it, followed by a truncation to the low energy subspace, followed by another similarity transformation. In Appendix~\ref{appalter}, we exhibit another effective Hamiltonians that are also constructed as similarity transformations which (partially) diagonalise followed by a truncation. We expect all such effective Hamiltonians to have this general structure, and differ from each other only through similarity transformations.
In particular, the Schrieffer--Wolff Effective Hamiltonian that we review in Appendix.~\ref{appalter} is hermitian by construction, which will make  the computation of time-evolution processes  conceptually clearer than  would be the case for a non-hermitian effective Hamiltonian.


\section{Application to UV divergent QFTs}
\label{sec:hampt}

Effective Hamiltonians have another use besides improving the estimates of the Hamiltonian Truncation spectrum. 
In this section, we show how effective Hamiltonians enable the systematic study of a wide variety of QFTs which have UV divergences using Hamiltonian Truncation.
We start by first reviewing the Hamiltonian Truncation set up that we shall use as well as the problem of UV divergences in this context. 

\subsection{Review of TCSA}

Hamiltonian Truncation is a generalisation of the Rayleigh--Ritz method, often used in quantum mechanics, to Quantum Field Theory. 
Several variations of this idea exist, which differ by placing the theory on different manifolds or truncating the Hilbert space in different quantisation schemes. For example, see  \cite{Anand:2020gnn}, the  reviews \cite{Fitzpatrick:2022dwq,Konik-review} and references within.
In the main body of this work, we focus on the Truncated Conformal Space Approach (TCSA)~\cite{Yurov:1989yu} and its higher dimensional generalisation~\cite{Hogervorst:2014rta}.

In the rest of this work, we analyse Hamiltonian Truncation in QFTs with $d\geq 2$ that are defined as relevant deformations away from an ultraviolet CFT. To regulate infrared divergences the CFT is placed on the ``cylinder" $\mathds{R}\times S_R^{d-1}$ where $R$ is the radius of the $d-1$ dimensional sphere. The Hamiltonian is given by
\be
H=H_\text{CFT} +V \quad , \quad \quad  \text{where}\quad \quad V= g R^{\Delta-d}\int_{S_R^{d-1}} d^{d-1}x \phi_\Delta(0,\vec x) \ , \label{tcsafirst}
\ee
where $g$ is a dimensionless coupling. For simplicity we will perturb $H_\text{CFT}$ using a single relevant primary operator, but generalisation to the case of multiple perturbing operators is possible. Thanks to the operator-state correspondence, the Hamiltonian on the cylinder is related to the  dilatation operator, and thus
the eigenvalues are given by the scaling dimensions of local operators
\be
\langle \calO_j | H_\text{CFT} | \calO_i \rangle = E_i\, \delta_{ij }  \quad , \quad \quad  \text{where}\quad \quad E_i = \Delta_i/R \  .
\ee
Next we truncate the Hamiltonian $H$ to the low energy sector spanned by the states $\{|\calO_i \rangle\}$ with scaling dimension less that the TCSA cutoff $\Delta_i\le\D$,
leading to the finite dimensional matrix
\be
H_{ij} \equiv  \Delta_i  \delta_{ij }+ V_{ij}\quad , \quad \quad  \Delta_i \leq \Delta_T \ , \label{Hraw}
\ee
where we have defined $V_{ij}\equiv R\langle \calO_j | V| \calO_i \rangle$. Finally, we are required to take the continuum field theory limit by diagonalising a sequence of matrices $H_{ij}$ with increasingly large values of $\Delta_T$ and then extrapolating the spectrum towards the limit $\Delta_T\rightarrow \infty$. This method of computing the spectrum is non-perturbative and valid  at strong coupling $g\gtrsim 1$. 

It  is often useful to compare the non-perturbative spectrum of Hamiltonian Truncation, at weak coupling $g \ll 1$, with the spectrum computed 
with the Rayleigh--Schr\"{o}dinger (RS) perturbation theory
\be
\Eps_{i}\, R  = \Delta_i+
\underbrace{  \phantom{  \bigg|}  V_{ii}      \phantom{  \bigg|}   }_{\Eps_i^{(1)}}+  
\underbrace{   \phantom{  \bigg|}  V_{ik} \frac{1}{\Delta_{ik}} V_{ki}    \phantom{  \bigg|}   }_{\Eps_i^{(2)}}+
\underbrace{   \phantom{  \bigg|} V_{ik} \frac{1}{\Delta_{ik}}  V_{kk^\prime}  \frac{1}{ \Delta_{ik^\prime}}  V_{k^\prime i}   - V_{ii} \, V_{ik}\frac{1}{\Delta_{ik}^2}   V_{ki} }_{\Eps_i^{(3)}}  +  \, O(V^4)\,, \label{pert1}
\ee
where $\Delta_{ij}= \Delta_i-\Delta_j$ and a sum over intermediate states $k\neq i$ is implicit. For future reference we recall that each order of perturbation theory can be computed in position space.
For instance the $n$-th order correction to the ground state, or Casimir energy, is given by the integrated  connected  $n$-point function of the perturbing operator 
\be
\Eps_{gs}^{(n)}  = -( - g)^n     S_{d-1}  /n! \int_{\mathds{R}^d} \prod_{i=1}^{n-1}{d^dx_i}  |x_i|^{\Delta-d} \, \langle  \phi_\Delta(x_1) \cdots \phi_\Delta(x_{n-1}) \phi_\Delta(1)   \rangle_c  \, , \label{pert2}
\ee
where $S_{d-1}=2 \pi^{\frac{d}{2}}/\Gamma(d/2)$ and the spacetime coordinate denoted as 1 represents a unit vector in $\mathds{R}^d$.
In \reef{pert2} we have set $R=1$, in the rest of the paper we  measure all dimensionful quantities in units of $R$.
Similar expressions can be derived for the excited states.  Further details can be found in \cite{EliasMiro:2021aof}.

In obtaining \reef{pert2} a Weyl transformation from the ``cylinder"  $ \mathds{R}\times S_R^{d-1}$ into the plane $\mathds{R}^d$ has been used. 
From here on we use $x_i$ and $\phi_\Delta(x_i)$ to denote coordinates and fields on $\mathds{R}^d$.  We distinguish fields on the cylinder  $\phi_\Delta(\tau_i ,\vec x_i)$    by using $(\tau, \vec x_i)$ coordinates, where   $\vec x_i$ is a vector in  $S^{d-1}$  and $\tau_i$ is the cylinder time coordinate.

A beautiful aspect of Hamiltonian Truncation is its potential as a universal method for computing strongly coupled renormalisation group flows away from any UV CFT. Nevertheless there are important challenges to overcome before this potential may be fully realised. In particular, it is not understood in all cases how renormalisation is to be implemented within TCSA, when the underlying QFT has UV divergences. When the deforming operator has dimension $\Delta \geq d/2$, the spectrum of the truncated  Hamiltonian \reef{Hraw} does not converge as $\Delta_T\rightarrow\infty$. The reason for this can be understood by employing RS perturbation theory. The 2nd order correction to the Casimir energy,
$
\Eps^{(2)}_0\propto \int_{\mathds{R}^d}  d^{d}x \langle \phi_\Delta( x) \phi_\Delta(1) \rangle  
$, 
diverges for  $2\Delta \geq d$  because  at $x\rightarrow 0$  the two-point function $\L\phi_\Delta(x) \phi_\Delta(0)\R =|x|^{-2\Delta}$ is singular and the measure has dimensions of $(\text{length})^d$.
Similarly, the integral of the connected three-point function is divergent
 for  $3\Delta \geq  2d$, but finite otherwise, and 
the integral of the connected four-point function is divergent only  for $4\Delta \geq  3d$.
The main point is simple:
as  the dimension of $\phi_\Delta$ is  increased further we encounter  UV divergences in higher order correlation-functions; 
until we reach marginality $\Delta=d$ where all $n$-point connected correlation functions  are UV divergent.

The presence of UV divergences is a common phenomena in QFT, and the way to deal with it is to carefully define the theory with appropriate counterterms and renormalisation conditions. In the context of Hamiltonian Truncation, it does not work in the same way and the renormalisation procedure requires a few extra steps that we explain next.

\subsection{How to Determine Counterterms in Hamiltonian Truncation?}

The Hamiltonian Truncation cutoff provides a natural regulator that removes UV divergences. The truncated theory is free of UV divergences because the maximal energy of the free-theory states is $E_i\leq\Delta_T/R$. This cutoff however is non-local because all $n$-particle states are bounded by the total energy $\Delta_T/R$ irrespective of the distance between the constituent particles. In this sense, Hamiltonian Truncation is very different from traditional local regulators, such as the short distance regulators (which cut out regions of position space where one or more operators get close), or the momentum cutoff regulator (which cuts the maximal momenta flowing in loop diagrams).

As a consequence of the breaking of locality that arises from the use of the $\Delta_T$ regulator, non-local counter-terms are needed for renormalisation in order to recover a local theory in the limit $\Delta_T\rightarrow\infty$. This was demonstrated in Refs.~\cite{Elias-Miro:2020qwz,Anand:2020qnp} where state-dependent counterterms (that were therefore non-local) were employed to renormalise the $\phi^4$ theory in $d=2+1$ dimensions.
Furthermore, the use of the $\Delta_T$-cutoff to regularise the theory introduces UV divergences at higher orders than would have been present had a local regulator been used: in Ref.~\cite{EliasMiro:2021aof} it was shown that for $\Delta\in[d/2+1/4 ,  \, 3d/4)$, the fourth order correction to the Casimir energy  is finite if a local regulator is employed but diverges if the $\Delta_T$ cutoff is used.

In this work we do not use $\Delta_T$ as a regulator of UV divergences. We instead implement renormalisation in a QFT with relevant but UV divergent deformations through the use of Effective Hamiltonians. 
In the rest of the section, we set out the general logical steps that we follow in our approach to renormalisation, before moving on to detailed computations.

\begin{enumerate}
\item[{\bf 1)}]  

In order to define a local QFT  with finite energy levels, we first renormalise all integrated local $n$-point functions  of the perturbing operator by employing a local regulator
\be
 \langle  \phi_\Delta(x_1)  \phi_\Delta(x_2)\cdots   \phi_\Delta(x_n) \rangle   \  \longrightarrow  \ \langle \prod_{i=1}^n \phi_\Delta(x_1)   \phi_\Delta(x_2) \cdots   \phi_\Delta(x_n) \rangle   \prod_{i < j} \theta(|x_i-x_j|-\epsilon)\,,
\ee
so that the integral $\int \prod_{i=1}^{n-1} d^dx_i |x_i|^{\Delta-d}$ of the last equation is finite. We then introduce all the  necessary  local counter-terms   
\begin{multline}
V(\eps)\equiv V+H_{c.t.}(\epsilon)= g R^{\Delta-d}\int\limits_{S_R^{d-1}} d^{d-1}x\,\phi_\Delta(0,\vec x) \\+ \sum_{\substack{2\le n\le M\\j}} g^n \delta(\eps)_{n,j} R^{\Delta_j-d}\int\limits_{S_R^{d-1}} d^{d-1}x\,\calO_j(0,\vec x) \,,
\end{multline}
where $M$ is some finite integer because the perturbation  $\phi_\Delta$ is relevant,  $\Delta < d$. As usual the addition of counterterms guarantees that the perturbative calculation of the spectrum is finite in the $\epsilon\rightarrow0$ for any order in perturbation theory. We are left with the following renormalised Hamiltonian
	\be
	H(\epsilon)_{ij} \equiv  \Delta_i  \delta_{ij }+ V(\epsilon)_{ij}\,, \label{Hrenone}
	\ee
	which we represent as an infinite dimensional matrix.
	
	\item[{ \bf 2)}] 
	
	Next, we compute the effective Hamiltonian with finite dimensionality that matches the low energy spectrum of \reef{Hrenone}. This matrix can be computed using perturbation theory to some fixed order in $g$. Schematically, it takes the form
	\be
	H_\text{eff}(\eps)_{ij}= H(\eps)_{ij} + \sum_{n\geq 2} H_\text{eff n}(\eps)_{ij} \quad , \quad \quad  \Delta_i \leq \Delta_T   \, ,
	\label{heffexp} 
	\ee
	where $H_\text{eff n}(\eps)$ is $O(g^n)$. 
	
	In this step, we work out as many terms in the expansion of $H_\text{eff}$ as we need to produce a matrix where all of its elements are finite in the limit $\epsilon\rightarrow0$, for fixed cutoff $\D$. Since there are only divergences as $\epsilon\rightarrow0$ up to a finite order $M$ in perturbation theory, which are absorbed into a finite number of counterterms, it will only be necessary to calculate the first $M$ orders of \reef{heffexp}.
	\item[{ \bf 3)}]
	
	Finally, we take the limit $\eps\rightarrow 0$ analytically. For future convenience we define the operator $K$ that arises in this limit
	\be
	H_\text{eff}= H_0+V + K \, .  \label{knot}
	\ee
\end{enumerate}

Once $O(g^M)$ terms are dropped from \reef{heffexp} in step {\bf 2)}, the effective Hamiltonian no longer has exactly the same low-energy spectrum as the original renormalised QFT. However, the (finite) neglected terms in $H_\text{eff}$ all have to vanish as $\D\rightarrow\infty$ to ensure that $H_\text{eff}$ matches the original renormalised Hamiltonian \reef{Hrenone} in this limit. As a result, estimates of the spectrum made using the three steps will ultimately tend towards the exact spectrum as $\D$ is increased. The vanishing of the higher order terms  can also be seen explicitly in \reef{hl2neat}; as $\D\rightarrow\infty$, $\Sigma_l\rightarrow \Sigma$ and $H_\text{eff}\rightarrow H_0+V$.

Another virtue of this construction is that it makes transparent the renormalisation scheme, which is fully specified in step {\bf 1)}. It can also be generalised by using alternative local regulators or Effective Hamiltonians, and can also be applied to different Hamiltonian Truncation approaches such as the massive Fock-space approaches \cite{Rychkov:2014eea,Elias-Miro:2017xxf} or lightcone conformal truncation \cite{Anand:2020gnn}.
For instance, in the massive Fock space approach, one could use a momentum cutoff regulator. 

Previously,  Refs.~\cite{Hogervorst:2014rta,Giokas:2011ix,Rychkov:2014eea,Rychkov:2015vap,Elias-Miro:2015bqk,Elias-Miro:2017xxf,Elias-Miro:2017tup,Cohen:2021erm} used Effective Hamiltonians to improve the convergence of the Hamiltonian Truncation spectrum as a function of the truncation cutoff. Most of these works studied the $\phi^4$-theory in $d<3$ spacetime dimensions, which is an ultraviolet finite theory.~\footnote{The only exception is Ref.~\cite{Giokas:2011ix}, which  studied RG flows of theories with UV divergencies using an Effective Hamiltonian. However, due to the value of the scaling dimension of the relevant deformation studied there, a second order Hamiltonian Truncation counterterm was enough to render the theory finite~\cite{EliasMiro:2021aof}. }
Effective Hamiltonians are also helpful to organise the Hamiltonian Truncation computations for perturbations that are  UV divergent. In particular in the next section  we show how to carry out Hamiltonian Truncation renormalisation at Leading Order (LO) and Next-to-leading order (NLO). We will also show in detail how the non-local UV divergences found in Ref.~\cite{EliasMiro:2021aof} are treated in this approach.

\section{Determining the Effective Hamiltonian}
\label{tcsaex}
\subsection{Leading Order}
\label{sec:secondo}

We begin with a calculation of an effective Hamiltonian at second order in perturbation theory.
 For simplicity, we choose our effective Hamiltonian to be the hermitian conjugate of the CFHL Hamiltonian, shown in \reef{heff2}. This $H_\text{eff 2}$ can be recast into the following integral over half the cylinder
\begin{align}
\left(H_\text{eff 2}\right)_{fi} &=-g^2S_{d-1}\int_{-\infty}^{0}d\tau\sum_{h\in{\cal H}_h} e^{\tau\left(\Delta_h-\Delta_i\right)}\int_{S_{d-1}} d^{d-1}x\matrixel{f}{\phi_\Delta(0,\vec{n})}{h}\matrixel{h}{\phi_\Delta(0,\vec{x})}{i}\, ,\label{eq:h2cyl}
\end{align}
where the notation $\phi_\Delta(0,\vec{n})$ indicates that the operator acts at zero cylinder time but at an arbitrary spatial coordinate in $S_{d-1}$. Since the sum is only over states in ${\cal H}_h$ above the HT cutoff, the quantity $\Delta_h-\Delta_i$ is always positive and the integral converges in the region $\tau\rightarrow-\infty$. By using the time evolution equation $\phi_\Delta(\tau,\vec{x})=e^{\tau H_0}\phi_\Delta(0,\vec{x})e^{-\tau H_0}$, and by Weyl transforming the integral from the cylinder $\mathds{R}\times S_{d-1}$ to the plane $\mathds{R}^d$, Eq.~\reef{eq:h2cyl} can be rewritten as 
\begin{align}
	\left(H_\text{eff 2}\right)_{fi}=-g^2S_{d-1}\int\limits_{\substack{0\le|x|<1}}d^dx|x|^{\Delta-d}\bra{f}{\phi_\Delta(1)\myline\phi_\Delta(x)}\ket{i}\,,\label{h2pl}
\end{align}
where the vertical dashed line indicates the insertion of a partial resolution of the identity 
\be
\myline\ \equiv \sum_{h\in\H_h}\ket{h}\bra{h} \, ,\label{dash}
\ee
which  depends on $\Delta_T$  through the definition of $\H_h$.

To evaluate the matrix element in \reef{h2pl}, we first calculate it ignoring the partial insertion of the identity, then expand the result in powers of $|x|$, and then only retain terms in which the power of $|x|$ exceeds the state dependent threshold $\Dp_{,i}\equiv\D-\Delta_i-\Delta$. The expansion is just the Weyl transformation of the expansion in powers of $e^\tau$ shown in \reef{eq:h2cyl}. The expansion converges for $0<|x|<1$ when the matrix element, given by the correlation function below 
\begin{equation}
	\matrixel{f}{\phi_\Delta(1)\phi_\Delta(x)}{i} \equiv\L\calO_f(\infty)\phi_\Delta(1)\phi_\Delta(x)\calO_i(0)\R\,,\label{eq:4ptcor}
\end{equation}
is finite. Here, we have used the notation $\calO(\infty)\equiv\lim_{s\rightarrow\infty}|s|^{2\Delta_{\calO}}\calO(s)$. This procedure for calculating correlation functions containing partial resolutions of the identity has been used in Ref.~\cite{Elias-Miro:2015bqk}, see also the review \cite{Konik-review}. 
Within the region $|1-x|<1$, the correlation function in \reef{eq:4ptcor} can be decomposed using
\begin{align}
	\L\calO_f(\infty)\phi_\Delta(1)\phi_\Delta(x)\calO_i(0)\R = \sum_{\calO}\L\calO_f(\infty)\calO(1)\calO_i(0)\R\,\L\calO(\infty)\phi_\Delta(1)\phi_\Delta(x)\R\,,\label{eq:decomp4pt}
\end{align}
where the sum runs over all scalar operators (primaries and descendants) in the CFT. This formula can also be obtained using the operator product expansion (OPE) for $\phi_\Delta(x)\phi_\Delta(1)$ inside the correlator. In appendix~\ref{app:ope}, the notation and derivation of \reef{eq:decomp4pt} is explained in greater detail along with generalisations of this formula which are useful for calculating effective Hamiltonians at higher orders in perturbation theory.  

The integral in \reef{h2pl} will be convergent in the region $x\rightarrow 0$ since \reef{eq:h2cyl} converges in the region $\tau\rightarrow-\infty$, due to the partial resolution of the identity. However, \reef{h2pl} may diverge in the region where $x\rightarrow 1$, if the three--point functions in the decomposition \reef{eq:decomp4pt} have a singularity there that is not integrable. This can be interpreted as an ultraviolet divergence in the underlying QFT to be regulated. In this case, we define the regulated effective Hamiltonian
\begin{align}
	\left(H_\text{eff 2}(\epsilon)\right)_{fi}=-g^2S_{d-1}\int\limits_{\substack{0\le|x|<1\\|1-x|>\epsilon}}d^dx|x|^{\Delta-d}\bra{f}{\phi_\Delta(1)\myline\phi_\Delta(x)}\ket{i}\,,\label{h2plreg}
\end{align}
where an $\epsilon$--ball has been cut out from the domain of integration at the site of the singularity, which acts as an ultraviolet regulator making the integral finite.

\begin{figure}[t]\centering
	\includegraphics[width=10cm]{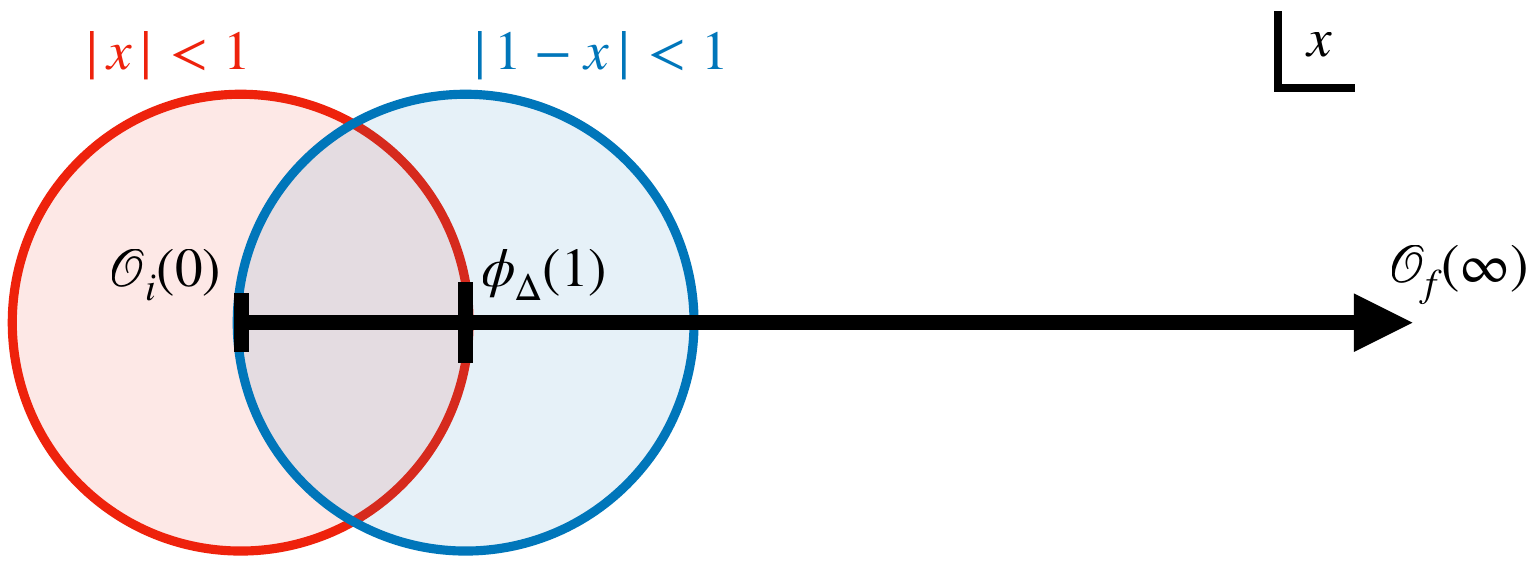} 
	\caption{Regions of integration in \reef{h2pl2reg}.} 
	\label{iregions}
\end{figure}

We now split the integral in \reef{h2plreg} into two integrals over complementary regions, and use \reef{eq:decomp4pt} to decompose the matrix element in the region where $|1-x|<1$, as shown below 
\begin{multline}
	\left(H_\text{eff 2}(\epsilon)\right)_{fi}=-g^2S_{d-1}\sum_{\calO}\matrixel{f}{\calO(1)}{i}\int\limits_{\substack{0\le|x|<1\\1>|1-x|>\epsilon}}d^dx|x|^{\Delta-d}\L\calO(\infty)\phi_\Delta(1)\myline{}^i\phi_\Delta(x)\R\\-g^2S_{d-1}\int\limits_{\substack{0\le|x|<1\\|1-x|\ge1}}d^dx|x|^{\Delta-d}\bra{f}{\phi_\Delta(1)\myline\phi_\Delta(x)}\ket{i}\,.\label{h2pl2reg}
\end{multline}
Since all the dependence on $x$ in each term of the sum over operators \reef{eq:decomp4pt} is contained within one of the three--point functions, the integral over $x$ and partial resolution of the identity should only be applied to those three--point functions, as indicated above. We have represented this partial resolution using the new symbol
\begin{align}
	\myline{}^i\equiv\sum_{\Delta_h>\D-\Delta_i}\ket{h}\bra{h}\,,\label{idi}
\end{align}
to emphasise that the restriction on which states are included in the sum retains dependence on the state $\ket{i}$, which it inherits from \reef{h2plreg}. See also Fig.~\ref{iregions} for an illustration of the regions of integration. 

The second integral in \reef{h2pl2reg} over the region $|1-x|>1$ is finite and has no dependence on the UV regulator scale $\epsilon$. It must therefore have a finite limit as $\D\rightarrow\infty$. Although it contains $\D$ dependent terms, it will not play a role in the renormalisation of UV divergences of $H_\text{eff}$. By contrast, the first term diverges in the $\epsilon\rightarrow0$ limit and requires renormalisation. For the rest of this section our focus will be on the first term.

In the sum over operators, in the first term of \reef{h2pl2reg}, the most divergent terms in the limit $\epsilon\rightarrow0$ will be terms in which $\calO(\infty)$ is a scalar primary. 
For simplicity from here on we assume that  in  the sum \reef{h2pl2reg}  
only  primary operators ${\cal O}$ give divergent contributions, but the derivation  can be generalised to include  divergent contributions from non-primary operators.  
In this case, the three--point function is constrained by conformal invariance to take the following series expansion
\begin{align}
	\L\calO(\infty)\phi_\Delta(1)\phi_\Delta(x)\R = \frac{f_{\calO\Delta\Delta}}{|1-x|^{2\Delta-\Delta_{\calO}}}=f_{\calO\Delta\Delta}\sum_{n=0}^\infty|x|^n\,C_n^{\Delta-\Delta_{\calO}/2}(\cos\theta)\,,\label{eq:3ptfunction}
\end{align}
where the $C^\alpha_n(\cos\theta)$ denote Gegenbauer polynomials. After accounting for the partial resolution of the identity by using the series expansion in $|x|$ and dropping terms with low powers we find that
\begin{multline}
	\left(H_\text{eff 2}(\epsilon)\right)_{fi}=-g^2S_{d-1}\sum_{\calO}\matrixel{f}{\calO(1)}{i}f_{\calO\Delta\Delta}\int\limits_{\substack{0\le|x|<1\\|1-x|>\epsilon}}d^dx|x|^{\Delta-d}\sum\limits^\infty_{n>\Dp_{,i}}|x|^nC_n^{\Delta-\Delta_{\calO}/2}(\cos\theta)\\
	+ \text{finite as }\epsilon\rightarrow0\,,\label{eq:heff2sumint}
\end{multline}
where the domain of integration has been extended to include the region $|1-x|>1$, which only introduces an $\epsilon$ independent correction and $\Dp_{,i}\equiv\D-\Delta_i-\Delta$. Using the fact that the integrals over Gegenbauer polynomials vanish for odd $n$ in the limit $\epsilon\rightarrow0$, and by introducing the notation
\begin{align}
	u_n^{\alpha,\,\epsilon}\equiv2(2n+\Delta)\int\limits_{\substack{0\le|x|<1\\|1-x|>\epsilon}}d^dx|x|^{2n+\Delta-d}C_{2n}^{\alpha}(\cos\theta)\,,\label{eq:uneps}
\end{align}
we simplify \reef{eq:heff2sumint} to read
\begin{align}
		\left(H_\text{eff 2}(\epsilon)\right)_{fi}=-\frac{g^2S_{d-1}}{2}\sum_{\calO}\matrixel{f}{\calO(1)}{i}f_{\calO\Delta\Delta}\sum^\infty_{2n>\Dp_{,i}}\frac{u_n^{\Delta-\Delta_\calO/2,\,\epsilon}}{2n+\Delta}+\text{finite}\dots  \label{lstep}
\end{align}
The sum over primaries is  finite for nonzero $\epsilon$, but can become divergent in the $\epsilon\rightarrow0$ limit if $2\Delta-\Delta_{\calO}\ge d$.
We note that the individual coefficients  $u_n^{\Delta-\Delta_{\cal O}/2,\eps}$ are always finite in this limit; however the infinite sum may still diverge. 	
	
After having completed this preliminary calculation we are ready to apply steps {\bf 1)} - {\bf 3)} described in the previous section. 
We first renormalise the theory by adding \emph{local} counterterms to the Hamiltonian as necessary, so that $H_\text{eff}$ is finite in the $\epsilon\rightarrow0$ limit
\begin{align}
	H_0+V \, \longrightarrow  \,  H_0+V+\sum\limits_{\substack{2\Delta-\Delta_{\calO}\ge d}}\lambda^\calO\int_{S_{d-1}}d^{d-1}x\,\calO(x)\,,\label{eq:localct}
\end{align}
where the couplings can be divided into a renormalised part plus a counterterm coupling $\lambda^\calO=\lambda^\calO_\text{ren}+\lambda^\calO_\text{ct}(\epsilon)$. A convenient choice of renormalisation scheme 
 is to define the counterterm coupling in the following way
\begin{align}
	\lambda^\calO_\text{ct}(\epsilon)&=g^2\int\limits_{\substack{0\le|x|<1\\1>|1-x|>\epsilon}}d^dx|x|^{\Delta-d}\L\calO(\infty)\phi_\Delta(1)\phi_\Delta(x)\R\,,\label{cterm2}\\
	& = \frac{g^2f_{\calO\Delta\Delta}}{2}\sum^\infty_{n=0}\frac{u_n^{\Delta-\Delta_\calO/2,\,\epsilon}}{2n+\Delta}\,.
\end{align}
With this choice, the $\epsilon\rightarrow0$ limit becomes well defined at leading order, i.e. $O(g^2)$. This completes the step {\bf 1)} at leading order.
Next we truncate the Hamiltonian, and compute the leading order Effective Hamiltonian for the renormalised theory in the right hand side of \reef{eq:localct} to complete step {\bf 2)}, and finally take the $\eps\rightarrow 0$ limit at the level of matrix elements to complete step {\bf 3)}.
By employing similar manipulations to the ones used to derive  \reef{lstep}, we find that the renormalised $H_\text{eff 2}$ in this limit is given by
\begin{multline}
	\left(H_\text{eff 2}\right)_{fi} = S_{d-1}\sum\limits_{\substack{2\Delta-\Delta_{\calO}-d>0}}\lambda^\calO_\text{ren}\matrixel{f}{\calO(1)}{i}\\+\frac{g^2S_{d-1}}{2}\sum\limits_{\substack{2\Delta-\Delta_{\calO}-d\ge0}}\matrixel{f}{\calO(1)}{i}f_{\calO\Delta\Delta}\sum^{2n\le\Dp_{,i}}_{n=0}\frac{u_n^{\Delta-\Delta_\calO/2}}{2n+\Delta}+\dots\label{eq:h2ren}
\end{multline}
In this expression, the sums over $n$  diverge in the $\D\rightarrow\infty$ limit. The $\dots$ represent other omitted terms that are finite in this limit. According  to the notation introduced in \reef{knot}, we identify last term above with
\be
K_2= S_{d-1}   \sum\limits_{\substack{2\Delta-\Delta_{\calO}-d\ge0}}
\matrixel{f}{\calO(1)}{i}  \left( 
\lambda^\calO_\text{ren}+\frac{g^2   f_{\calO\Delta\Delta}}{2}\sum^{2n\le\Dp_{,i}}_{n=0}\frac{u_n^{\Delta-\Delta_\calO/2}}{2n+\Delta} \right)+\cdots
\ee

Although the counterterm we introduced in \reef{eq:localct} was manifestly local, the renormalised effective Hamiltonian we are left with in \reef{eq:h2ren} has non-local interactions. The non-locality appears through the dependence of the sum over $n$ on $\Delta_i$ -- the scaling dimension of the specific initial state $\ket{i}$. Finally, we note that the $u_n^{\Delta,\,\epsilon}$ coefficients have a convenient expression in the $\epsilon\rightarrow0$ limit that can be derived from \reef{eq:uneps}
\begin{align}
	u_n^\Delta\equiv\lim_{\epsilon\rightarrow 0}u_n^{\Delta,\,\epsilon} &= 2S_{d-2}\int_{0}^{\pi}d\theta\,(\sin\theta)^{d-2}\,C^\Delta_{2n}(\cos\theta)\,, \\
	&=2S_{d-2}
	\sqrt{\pi }\, \Gamma \left(\tfrac{d-1}{2}\right) 
	\frac{\Gamma (n+\Delta )}{\Gamma (\Delta )n!} \, 
	\frac{\Gamma\left(n+\Delta-\frac{d-2}{2}\right)}{\Gamma \left(\Delta-\frac{d-2}{2} \right)\Gamma \left(n+\frac{d}{2}\right) }
	\, .\label{eq:un}
\end{align}

\subsection{Next-to-Leading Order}
\label{sec:thirdo}

We now turn to determining the dependence of $H_{\text{eff}\,3}$ on the cutoff $\D$. We do this by representing the effective Hamiltonian as a sum of position space integrals over CFT correlation functions and analysing those integrals' UV divergences. We take $H_{\text{eff}\,3}$ to be the hermitian conjugate of the third-order CFHL Hamiltonian shown in \reef{heff3}. Using the definition of $V$ in terms of a CFT operator in (\ref{tcsafirst},\ref{Hraw}), we rewrite $H_{\text{eff}\,3}$ as
\begin{multline}
	\left(H_{\text{eff}\,3}\right)_{fi}=g^3S_{d-1}\prod_{i=1}^{2}\int_Td\tau_i\int_{S_{d-1}}d^{d-1}x_i\\ \Bigg[\sum_{h,\,h^\prime\in\H_h}e^{\tau_2\Delta_{h^\prime h}}\,e^{\tau_1\Delta_{hi}}\matrixel{f}{\phi_{0,\vec{n}}}{h^\prime}\matrixel{h^\prime}{\phi_{0,\vec{x_2}}}{h}\matrixel{h}{\phi_{0,\vec{x_1}}}{i}\\ - 
	\sum_{l\in\H_l,\,h\in\H_h}e^{\tau_2\Delta_{li}}\,e^{\tau_1\Delta_{hl}}\matrixel{f}{\phi_{0,\vec{n}}}{h}\matrixel{h}{\phi_{0,\vec{x_1}}}{l}\matrixel{l}{\phi_{0,\vec{x_2}}}{i}\Bigg]\,,\label{h3cyl}
\end{multline}
where $T$ indicates the time ordered domain of integration $-\infty<\tau_1\le\tau_2\le0$. We represent differences in scaling dimension using $\Delta_{ij}\equiv\Delta_i-\Delta_j$, and we introduce the more compact notation $\phi_{0,\vec{x}}\equiv\phi_\Delta(0,\vec{x})$. Removing the exponential factors using the Heisenberg equation for the operators $\phi_\Delta(\tau,\vec{x})=e^{\tau H}\phi_\Delta(0,\vec{x})e^{-\tau H}$, and converting the integral from the cylinder to the plane yields the expression
\begin{multline}
	\left(H_{\text{eff}\,3}\right)_{fi}=g^3S_{d-1}\int_R\prod_{i=1}^2d^dx_i|x_i|^{\Delta-d}\Big[\bra{f}\phi_\Delta(1)\myline\phi_\Delta(x_2)\myline\phi_\Delta(x_1)\ket{i}\\-\sum_{l\in\H_l}\bra{f}\phi_\Delta(1)\myline\phi_\Delta(x_1)\ket{l}\matrixel{l}{\phi_\Delta(x_2)}{i}\Big]\,.\label{h3plane}
\end{multline}
Here, $R$ indicates the radially ordered integration domain $0<|x_1|\le|x_2|\le1$ and the dashed vertical lines indicate the sum over high energy states shown in \reef{dash}. The integrals above have no divergences from the $|x_i|\rightarrow0$ regions (which correspond to infrared divergences from the $\tau_i\rightarrow -\infty$ regions on the cylinder) because low energy states have been excluded from the sums.

The integrals in \reef{h3plane} can have UV divergences, coming from the regions $x_i\rightarrow1$ and $x_1\rightarrow x_2$. As in section~\ref{sec:secondo}, we regulate them by cutting out $\epsilon$-balls from the integration domain around these singular points. Furthermore, the local counterterms that we added to the theory at second order make an extra $\epsilon$ dependent contribution to $H_{\text{eff}}$ that is $O(g^3)$. Accounting for the local regulator dependence leads to the following expression
\begin{multline}
	\left(H_{\text{eff}\,3}(\epsilon)\right)_{fi}=g^3S_{d-1}\int\limits_{\substack{R,\;|1-x_i|>\epsilon,\\|x_2-x_1|>\epsilon|x_2|}}\prod_{i=1}^2d^dx_i|x_i|^{\Delta-d}\Big[\bra{f}\phi_\Delta(1)\myline\phi_\Delta(x_2)\myline\phi_\Delta(x_1)\ket{i}\\-\sum_{l\in\H_l}\bra{f}\phi_\Delta(1)\myline\phi_\Delta(x_1)\ket{l}\matrixel{l}{\phi_\Delta(x_2)}{i}\Big]\\
	-gS_{d-1}\sum_{\Delta_{\calO}<2\Delta-d}\lambda^\calO_\text{ct}(\epsilon)\int\limits_{\substack{0<|x|\le1,\\|1-x|>\epsilon}} d^dx|x|^{\Delta-d}\Big[\bra{f}\calO(1)\myline\phi_\Delta(x)\ket{i}\\+|x|^{\Delta_{\calO}-\Delta}\bra{f}\phi_\Delta(1)\myline\calO(x)\ket{i}\Big]\,,\label{h3planeeps}
\end{multline}
where $\lambda^\calO_\text{ct}(\epsilon)$ are the counterterms introduced in \reef{cterm2}. To preserve $\tau$ translation invariance on the cylinder, the sizes of regions cut out must be independent of the $\tau$ coordinate. After mapping to the plane, the $\epsilon$-ball widths then have to scale proportional to the radial coordinate. For this reason, the $\epsilon$-ball regulating the $x_1\rightarrow x_2$ singularity has a width that depends on the spatial coordinates like $|x_2-x_1|>\epsilon|x_2|$  (recall $d s^2_{\mathds{R}^d}= r^2  d s^2_\text{cyl}$).

As a result of cancellations between the first term and subtracted terms in \reef{h3planeeps}, the only divergence as $\epsilon\rightarrow0$ comes from the region of the first integral where all three fields are approaching each other (both $x_1\rightarrow1$ and $x_2\rightarrow1$ together). For example, the subdivergence coming from the region of the first integral where $x_2\rightarrow1$ with $x_1$ far away is canceled by the first term proportional to $\lambda^\calO_\text{ct}(\epsilon)$. This can be seen after inputting the expression \reef{cterm2}.

\begin{figure}\centering
	\includegraphics[width=11cm]{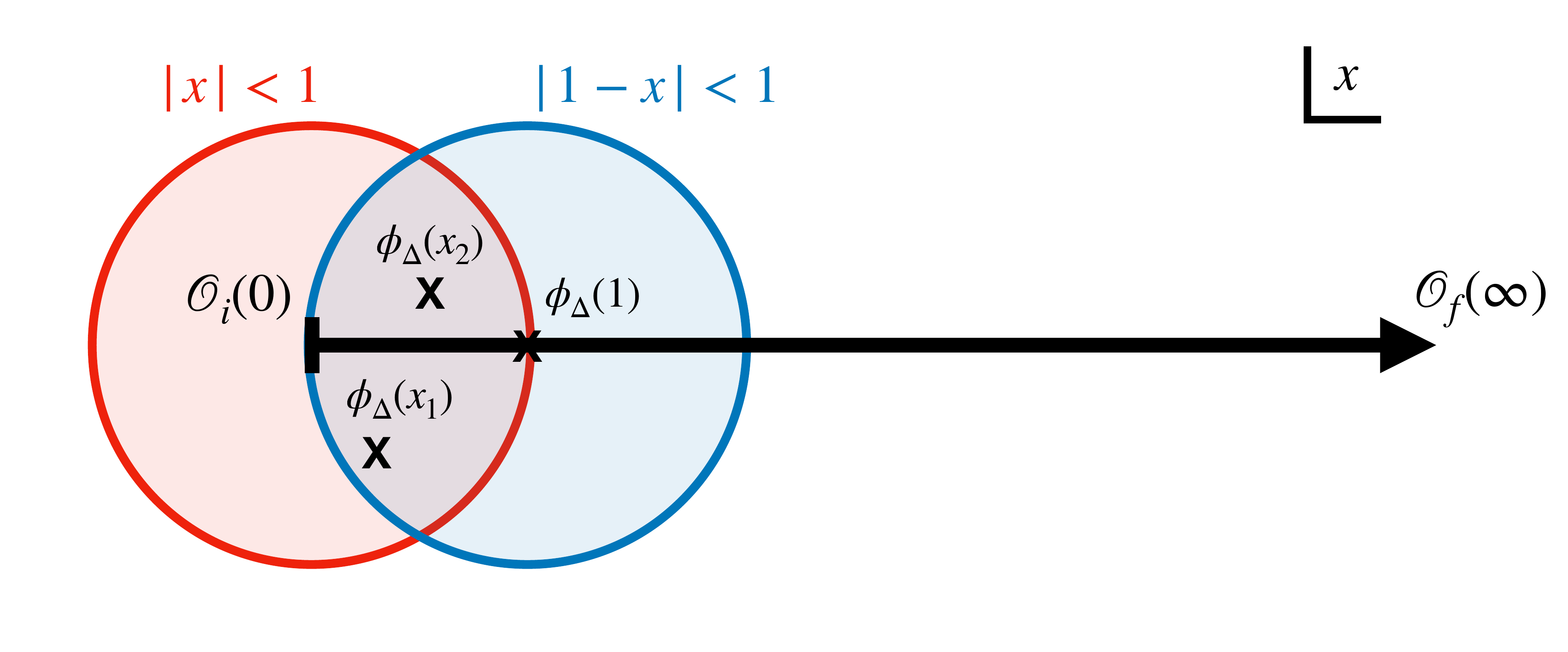} 
	\caption{The region where points are radially ordered so that $0<|x_1|\le|x_2|\le 1$ and satisfy $|1-x_i|\le1$. In this region, \reef{factor} may be used to simplify matrix elements.\label{fig:regh3}  } 
\end{figure}

Since the only parts of the expression \reef{h3planeeps} which can potentially diverge in the limit $\D\rightarrow\infty$ are also the parts that diverge in the $\epsilon\rightarrow0$ limit, we focus our attention on the region of the integrals where the points are configured as indicated in Fig.~\ref{fig:regh3} (the region where the two circles intersect). In this region, we can simplify matrix elements using
\begin{align}
	\matrixel{f}{\phi_\Delta(1)\phi_\Delta(x_2)\phi_\Delta(x_1)}{i} = \sum_{\calO}\matrixel{f}{\calO(1)}{i}\L\calO(\infty)\phi_\Delta(1)\phi_\Delta(x_2)\phi_\Delta(x_1)\R\,,\label{factor}
\end{align}
which can be regarded as a generalisation of the OPE to the case where three fields (rather than two) are approaching one another. More details are provided in Appendix~\ref{app:ope}. 

The term in \reef{factor} which gives the strongest singularity when the three $\phi_\Delta$ fields come together is the term with $\calO=\mathds{1}$. It therefore makes the contribution to $H_\text{eff 3}$ that is most UV divergent. For simplicity, we will only explicitly work out this contribution. We find that
\begin{multline}
	\left(H_{\text{eff}\,3}(\epsilon)\right)_{fi}\supset g^3S_{d-1}\delta_{fi}\int\limits_{\substack{R,\;|1-x_i|>\epsilon,\\|x_2-x_1|>\epsilon|x_2|}}\prod_{i=1}^2d^dx_i|x_i|^{\Delta-d}\Big[\L\phi_\Delta(1)\myline{}^i\phi_\Delta(x_2)\myline{}^i\phi_\Delta(x_1)\R\\-\sum_{\Delta_l\le\D-\Delta_i}\bra{0}\phi_\Delta(1)\myline{}^i\phi_\Delta(x_1)\ket{l}\bra{l}\phi_\Delta(x_2)\ket{0}\Big]\,,\label{h3id}
\end{multline}
where the symbol $\myline{}^i$ represents the insertion of a partial resolution of the identity, where the cutoff between high and low energy states depends on the scaling dimension of state $i$, as in \reef{idi}. Similarly, the cutoff on the sum over low energy states in \reef{h3id} retains dependence on $\Delta_i$. There are interesting QFTs, including $\phi^4$ theory in $d=2+1$ and minimal models deformed by the $\phi_{1,3}$ operator in $d=1+1$, for which \reef{h3id} is the only contribution to $H_\text{eff 3}$ that diverges in the $\epsilon\rightarrow0$ limit.

To evaluate the top line of \reef{h3id}, we take the three point function and express it as a series expansion in powers of $|x_2|$ and $|x_1|/|x_2|$ using
\begin{align}
	\L\phi_\Delta(1)\phi_\Delta(x_2)\phi_\Delta(x_1)\R&=\frac{f_{\Delta\Delta\Delta}}{|1-x_2|^\Delta|x_2-x_1|^\Delta|1-x_1|^\Delta}\,,\\
	&=\frac{f_{\Delta\Delta\Delta}}{|x_2|^\Delta|x_1|^\Delta}\sum\limits_{n_1,n_2,n_3=0}^{\infty}|x_2|^{n_1+n_2+\Delta}\left(\frac{|x_1|}{|x_2|}\right)^{n_1+n_3+\Delta}\nonumber\\&\qquad\qquad\quad\;\;\;\times C^{\Delta/2}_{n_1}(\cos\theta_1)C^{\Delta/2}_{n_2}(\cos\theta_2)C^{\Delta/2}_{n_3}(\cos\gamma)\,,\label{3ptseries}
\end{align}
where $\gamma$ represents the angle between points $x_1$ and $x_2$. On the cylinder, this corresponds to writing the three point function as a series expansion in $e^{\tau_2}$ and $e^{\tau_1-\tau_2}$. In \reef{h3cyl}, we see that in each term of the expansion, $e^{\tau_2}$ and $e^{\tau_1-\tau_2}$ are raised to the power $\Delta_h-\Delta_i$ for some $\Delta_h>\D$. By comparing \reef{3ptseries} with \reef{h3cyl}, we find that only terms satisfying the condition below should be included when evaluating the first line of $H_\text{eff 3}(\epsilon)$
\begin{align*}
	n_1+n_2&>\D-\Delta-\Delta_i\equiv\Dp_{,i}\,,\\
	n_1+n_3&>\Dp_{,i}\,.
\end{align*}

To evaluate the second line of \reef{h3id}, we instead expand the three point function in powers of $|x_1|$ and $|x_2|/|x_1|$. This is because on the cylinder in \reef{h3cyl}, the second line may be expressed as a series expansion in $e^{\tau_1}$ and $e^{\tau_2-\tau_1}$. In each term, these exponentials are raised to the powers $\Delta_h-\Delta_i$ and $\Delta_l-\Delta_i$. The series expansion that we need for the three point function is also given by \reef{3ptseries}, except with $|x_1|$ swapped with $|x_2|$. We find that only terms satisfying the condition below should be included when evaluating the second line of $H_\text{eff 3}(\epsilon)$
\begin{align*}
	n_1+n_2&>\Dp_{,i}\,,\\
	n_1+n_3&\le\Dp_{,i}\,.
\end{align*}

When plugging the series expansions for the three point function into \reef{h3id}, we encounter integrals of the form below, which we represent using the function $X^\epsilon_{n_1,n_2,n_3}$
\begin{align}
	\frac{X^\epsilon_{n_1,n_2,n_3}}{\left(n_1+n_2+\Delta\right)\left(n_1+n_3+\Delta\right)}\equiv&\int\limits_{\substack{R,\;|1-x_i|>\epsilon,\\|x_2-x_1|>\epsilon|x_2|}}\prod_{i=1}^2\frac{d^dx_i}{|x_i|^d}\,\,|x_2|^{n_1+n_2+\Delta}\left(\frac{|x_1|}{|x_2|}\right)^{n_1+n_3+\Delta}\nonumber\\&\qquad\times C^{\Delta/2}_{n_1}(\cos\theta_1)C^{\Delta/2}_{n_2}(\cos\theta_2)C^{\Delta/2}_{n_3}(\cos\gamma)\,.\label{xfn}
\end{align}
In the $\epsilon\rightarrow0$ limit, the expression above is finite, and can be simplified further 
\begin{align}
	X_{n_1,n_2,n_3}\equiv\lim\limits_{\epsilon\rightarrow 0}X^\epsilon_{n_1,n_2,n_3}=\int_{S_{d-1}}\prod_{i=1}^2d^{d-1}x_i\;C^{\Delta/2}_{n_1}(\cos\theta_1)C^{\Delta/2}_{n_2}(\cos\theta_2)C^{\Delta/2}_{n_3}(\cos\gamma)\,.\label{xlim}
\end{align}
In appendix~\ref{sec:formulae}, we evaluate the integrals above to derive an alternative expression for $X_{n_1,n_2,n_3}$, which can be evaluated numerically more easily. We show that this function is symmetric in all its $n_i$ indices, and vanishes unless the $n_i$ are all even, or all odd.

After combining \reef{h3id} with \reef{3ptseries} and \reef{xfn}, and truncating to retain only the terms we need, we are left with the following expression
\begin{multline}
	\left(H_{\text{eff}\,3}(\epsilon)\right)_{fi}\supset g^3S_{d-1}\delta_{fi}f_{\Delta\Delta\Delta}\sum\limits^\infty_{\substack{n_1+n_2>\Dp_{,i}\\n_1+n_3>\Dp_{,i}}}\frac{X^\epsilon_{n_1,n_2,n_3}}{\left(n_1+n_2+\Delta\right)\left(n_1+n_3+\Delta\right)} \\ -g^3S_{d-1}\delta_{fi}f_{\Delta\Delta\Delta}\sum\limits^{n_1+n_3\le\Dp_{,i}}_{\substack{n_1+n_2>\Dp_{,i}}}\frac{X^\epsilon_{n_1,n_2,n_3}}{\left(n_1+n_2+\Delta\right)\left(n_2-n_3\right)}\,.\label{h3x}
\end{multline}
Although the $X^\epsilon_{n_1,n_2,n_3}$ are each finite in the limit $\epsilon\rightarrow0$, the infinite sums in \reef{h3x} diverge when $3\Delta\ge2d$. In this case, we renormalise the theory by adding a local counterterm proportional to the identity operator
\begin{equation}
	H_0+V\longrightarrow H_0+V+\lambda^{\mathds{1}}\int_{S_{d-1}}d^{d-1}x\;\mathds{1}\,.
\end{equation}
It is convenient to pick a scheme in which the counterterm is just the full integrated three point function
\begin{align}
	\lambda^\mathds{1}_\text{ct}(\epsilon) & = -g^3\int\limits_{\substack{R,\;|1-x_i|>\epsilon,\\|x_2-x_1|>\epsilon|x_2|}}\prod_{i=1}^2d^dx_i|x_i|^{\Delta-d}\L\phi_\Delta(1)\phi_\Delta(x_2)\phi_\Delta(x_1)\R\,,\\
	&=-g^3f_{\Delta\Delta\Delta}\sum\limits^\infty_{n_1,n_2,n_3=0}\frac{X^\epsilon_{n_1,n_2,n_3}}{\left(n_1+n_2+\Delta\right)\left(n_1+n_3+\Delta\right)}\,.\label{ct3sum}
\end{align}

The matrix $K$ defined in \reef{knot} is the effective hamiltonian with counterterms subtracted in the $\epsilon\rightarrow0$ limit. Its contribution from \reef{h3x} and \reef{ct3sum} is given by
\begin{multline}
	\left(K_3\right)_{fi}=-g^3S_{d-1}\delta_{fi}f_{\Delta\Delta\Delta}\sum\limits^{\substack{n_1+n_2\le\Dp_{,i}\\n_1+n_3\le\Dp_{,i}}}_{n_1,n_2,n_3=0}\frac{X_{n_1,n_2,n_3}}{\left(n_1+n_2+\Delta\right)\left(n_1+n_3+\Delta\right)}\\
	-g^3S_{d-1}\delta_{fi}f_{\Delta\Delta\Delta}\sum\limits^{n_1+n_3\le\Dp_{,i}}_{\substack{n_1+n_2>\Dp_{,i}}}\Bigg[\frac{X_{n_1,n_2,n_3}}{\left(n_1+n_2+\Delta\right)\left(n_2-n_3\right)}\\+\frac{2X_{n_1,n_2,n_3}}{\left(n_1+n_2+\Delta\right)\left(n_1+n_3+\Delta\right)}\Bigg]\,.\label{K3}
\end{multline}
In the first line, the sums only run over contributions from exchanged states with low scaling dimension. The second line includes mixed sums, which account for contributions from states exchanged with scaling dimensions both below and above $\Dp_{,i}$. For any fixed value of the cutoff, the elements of $K_3$ will all be finite, however in the $\D\rightarrow\infty$ limit they can diverge.

When $3\Delta-2d\ge0$ a contribution to $K_2$ that we ignored in Section~\ref{sec:secondo} can become important. The contribution itself vanishes in the $\D\rightarrow\infty$ limit, however it makes a correction to energy eigenvalues at third order through
\begin{align}
	\delta \calE_i \,R = \sum_{l\in\H_l}\left(\frac{V_{il}(\delta K_2)_{li}}{\Delta_{il}}+\frac{(\delta K_2)_{il}V_{li}}{\Delta_{il}}\right)\,,\label{deltaKcorr}
\end{align}
which actually is divergent. To get a finite spectrum, this contribution must be included. It comes from taking $\calO=\phi_\Delta$ in \reef{lstep}, and is given by
\begin{equation}
	(\delta K_2)_{fi} = -\frac{g^2S_{d-1}}{2}\matrixel{f}{\phi_\Delta(1)}{i}f_{\Delta\Delta\Delta}\sum^\infty_{2n>\Dp_{,i}}\frac{u_n^{\Delta/2}}{2n+\Delta}\,.
\end{equation}
The shift in energy eigenvalues induced by the above through \reef{deltaKcorr} cancels the shift coming from the second line of $K_3$ in \reef{K3} in the $\D\rightarrow\infty$ limit.

\section{Locality of the Effective Hamiltonian $H_\text{eff}$}
\label{sec:cutoffdep}

The Effective Hamiltonian $H_\text{eff}$ has two features which make it appear non-local. Firstly, it contains interactions which cannot be expressed as integrated local operator densities such as \reef{eq:h2ren}, and secondly, it acts on a truncated Hilbert space that is spanned by states with total energy less than $\D/R$. This is a non-local modification of the QFT, since the maximum energy of excitations in one region of space now depends on the energy carried by fields in distant regions of space. However, $H_\text{eff}$ is constructed so that its spectrum matches the local QFT, so these non-local effects cancel one another when calculating observables.

In this section, we perform a check for consistency between $H_\text{eff}$ and the locally regulated QFT. Specifically, we check the fourth order perturbative correction to the ground state energy, which in Rayleigh--Schr{\"o}dinger perturbation theory 
 is given by
\begin{align}
	\calE_0^{(4)}R=-\frac{V_{0k}V_{kk^\prime}V_{k^\prime k^{\prime\prime}}V_{k^{\prime\prime}0}}{\Delta_k\Delta_{k^\prime}\Delta_{k^{\prime\prime}}}+\frac{V_{0k}V_{k0}}{\Delta_k}\cdot\frac{V_{0k^\prime}V_{k^\prime0}}{\Delta^2_{k^\prime}}\,,
	\label{eq:e4}
\end{align}
where the sums run over CFT states $k,\,k^\prime,\,k^{\prime\prime}\neq0$ and we have used $V_{00}=0$. In raw TCSA, these sums should all be cut off so that $\Delta_k\le\D$. 

We now demonstrate that $\calE_0^{(4)}$ is UV finite in $H_\text{eff}$, when it is also finite in the local theory. We focus on the fourth order, as it is the lowest order in perturbation theory for which there is a subtraction term, and cancellations between the leading and subtraction terms may be spoilt by the UV regulator. This quantity was found to diverge in the limit $\D\rightarrow\infty$ in raw TCSA~\cite{EliasMiro:2021aof}, even when the corresponding quantity in the locally regulated theory was finite.
Indeed, for large $\D$, conformal symmetry constrains \reef{eq:e4} to grow as \cite{EliasMiro:2021aof}
\begin{align}
	\calE_0^{(4)}\sim-\frac{g^4S^2_{d-1}}{4R}\left\{\sum_{n_1,\,n_2=0}^{2n_1+2n_2+2\Delta\le\D}-\sum_{n_1,\,n_2=0}^{\substack{2n_1+\Delta\le\D \\ 2n_2+\Delta\le\D}}\right\}\frac{u_{n_1}^\Delta}{(2n_1+\Delta)^2}\cdot\frac{u^\Delta_{n_2}}{2n_2+\Delta}\,,
	\label{eq:sumdiff}
\end{align}
where the $u^\Delta_n$ are given by \reef{eq:un}. This difference between sums diverges in the limit $\D\rightarrow\infty$ for $\Delta\ge d/2+1/4$. 

When using a local regulator, it is more convenient to use the conformal perturbation theory expression for the coefficient
\be
	\calE_0^{(4)}= -\frac{g^4S_{d-1}}{4!R} \int \prod_{i=1}^3 d^dx_i   \left|x_i\right|^{\Delta-d} 
	\langle \phi_{x_1} \phi_{x_2}\phi_{x_3}\phi_{1}\rangle_c \, . \label{c4}
\ee
This expression is finite (in the limit that the local regulator is removed) provided that $\Delta<3d/4$, in contrast with raw TCSA.

So far though, our analysis has not accounted for any extra contributions to $H_\text{eff}$. To remove divergences in $\D$ at the fourth order in perturbation theory, we should in principle consider contributions to $H_\text{eff}$ up to fourth order in the coupling. Nevertheless, we find that only $H_\text{eff 2}$ provides the contribution to $\calE_0^{(4)}$ needed to restore consistency with the local theory.

The expression for $H_\text{eff 2}$ as a sum over operators $\calO$ is provided in \reef{eq:h2ren}. For a deformation with $\Delta$ just above $d/2+1/4$, the term in $H_\text{eff 2}$ with $\calO=\mathds{1}$ must always be retained. Other operators must also be retained if they are sufficiently relevant, but since their presence is model dependent, they cannot cancel the universal divergence appearing in \reef{eq:sumdiff}. We therefore consider the contribution to $\calE_0^{(4)}$ coming from the interaction
\begin{align}
	\left(K_2\right)_{fi}=\frac{g^2S_{d-1}}{2}\delta_{fi}\sum_{n=0}^{2n+\Delta+\Delta_i\le\D}\frac{u_n^\Delta}{2n+\Delta}\,.\label{eq:h2simple}
\end{align}
Since \reef{eq:h2simple} is a diagonal matrix, the only contribution it can make to $\calE_0^{(4)}$ is given by
\begin{align}
	\delta\calE_0^{(4)}R=\frac{V_{0k}\left(K_2\right)_{kk^\prime}V_{k^\prime 0}}{\Delta_k\Delta_{k^{\prime}}}   
	 -   \left(K_2\right)_{00}\frac{V_{0k}V_{k0}}{\Delta_k^2}\,.\label{eq:h24}
\end{align}
The last factor in  \reef{eq:h24} is given by
\begin{align}
	\frac{V_{0k}V_{k0}}{\Delta_k^2}=\frac{g^2S_{d-1}}{2}\sum_{n=0}^{2n+\Delta\le\D}\frac{u_n^\Delta}{(2n+\Delta)^2}\, .
\end{align}
We use this result to rewrite both terms in \reef{eq:h24}, finding that the first term equals minus the first sum in \reef{eq:sumdiff}, and the second term equals minus the second sum in \reef{eq:sumdiff}. We conclude that \reef{eq:h24} and \reef{eq:sumdiff} cancel each other exactly.

 We have thus explicitly shown that a non-local interaction in $H_\text{eff 2}$ cancels a UV divergence that arises at higher order in perturbation theory, due to the non-local nature of the Hamiltonian Truncation cutoff $\D$. The finiteness of $\calE_0^{(4)}$ for $\Delta<3d/4$ is necessary for consistency with the locally regulated theory.

Furthermore, we can see that $H_\text{eff 3}$ or $H_\text{eff 4}$ will not make a contribution to the spectrum that diverges in the $\D\rightarrow\infty$ limit, unless $\Delta\ge2d/3$ so that a local counterterm beyond second order is needed to renormalise the theory. By assumption, the infinite dimensional matrix $H(\epsilon)=\Delta+V(\epsilon)$ incorporates all the local counterterms required for a finite spectrum. If $\Delta<2d/3$, then $H(\epsilon)$ will have no terms that depend on the coupling as $g^3$ or $g^4$. In the large $\D$ limit, it can be seen that $H_\text{eff}$ coincides with $H(\epsilon)$, for instance from \reef{hl2neat}. We therefore expect $H_\text{eff 3}$ and $H_\text{eff 4}$ to vanish in the $\D\rightarrow\infty$ limit.

In appendix~\ref{appv6} we show that the finiteness of the spectrum proceeds to $O(V^6)$ as expected, thanks to  the addition of the non-local $K_2$.
We also comment on the connections of this calculation with previous studies in $\phi^4$ theory.

\section{Discussion}
\label{sec:disc}

In this work, we have described a procedure for applying Hamiltonian Truncation when the QFT under investigation has UV divergences that require renormalisation. Our procedure involves extra steps beyond those required for renormalisation in perturbative QFT with a local regulator, such as dim-reg. This is because Hamiltonian Truncation calculations involve approximating the original QFT with a quantum theory in a finite dimensional Hilbert space, which is a non-local modification of the original theory. We believe this is an important step towards turning Hamiltonian truncation into a universal tool, that can be applied to any renormalisable QFT.
	
	To implement renormalisation, we first had to define an Effective Hamiltonian that shared the exact low energy spectrum of the original Hamiltonian but had finite dimensionality. We provided an alternative representation for an effective Hamiltonian, first defined in \cite{Cohen:2021erm} in Section~\ref{sec:heff},
	and proved that the spectra of the UV theory and the Effective Hamiltonian match. 
	
	Our main analysis includes a perturbative calculation of this Effective Hamiltonian, applicable to QFTs defined as an ultraviolet CFT deformed with a relevant operator. By first introducing a local regulator and local counterterms for the QFT, and then defining an effective Hamiltonian for this finite theory that allows the local regulator to be removed analytically, we demonstrated how ultraviolet divergences can be consistently canceled, and how renormalisation can be applied in the context of Hamiltonian Truncation. 
	
	Our renormalisation procedure makes the choice of scheme transparent, enabling direct comparison between Hamiltonian Truncation calculations and results derived using other methods. It also provides explicit expressions for the non-local interactions that must be included in the renormalised effective Hamiltonian to ensure UV finiteness. As well as presenting a general argument, valid at all orders in perturbation theory for the consistency of our approach, we have also provided direct non-trivial checks using perturbation theory up to $O(V^6)$, showing the cancellation of UV divergences.
	
	It is important now to apply this procedure non-perturbatively to numerical Hamiltonian Truncation calculations in QFTs with UV divergences. For example, these methods are well suited to studying conjectured renormalisation group flows between two-dimensional CFTs, e.g.~\cite{Gaiotto:2012np,Fei:2014xta,Klebanov:2022syt}, as well as exploring the rich variety of strongly coupled flows in $d=1+1$ more generally. It will also be very worthwhile to investigate QFTs in higher dimensions, such as $\phi^4$-theory in $d=2+1$ (building on an earlier exploratory study \cite{Elias-Miro:2020qwz} using the massive Fock space approach), and Quantum Electrodynamics in $d=2+1$.

\section*{Acknowledgements}

We are grateful to  Kara Farnsworth, Liam Fitzpatrick, Ami Katz, Markus Luty, and Matt Walters for interesting discussions.
This project has received funding from the 
European Research Council, grant agreement n.~101039756.

\appendix


\section{Alternative Effective Hamiltonians}

\label{appalter}

\subsection{The Schrieffer--Wolff Effective Hamiltonian}
\label{sec:sw}

The Schrieffer--Wolff Hamiltonian \cite{PhysRev.149.491} may be interpreted as an effective Hamiltonian (see also \cite{coleman2015introduction} for a pedagogical introduction). It is defined as the following canonical transformation and projection of the full theory Hamiltonian:
\begin{align}
	H^{SW}_\text{eff} & = \left[e^S\left(H_0+g V\right)e^{-S}\right]_l\, , \label{eq:swdef}
\end{align}
where $g$ is the coupling parameter, and the subscript $l$ indicates projection of the operator in square brackets onto the subspace ${\cal H}_l$. Then $H^{SW}_\text{eff}$ is fully specified if we further require that:
\begin{enumerate}
	\item $S$ is analytic in the coupling $g$ and vanishes when $g=0$ so that
	\begin{align}
		S = g S_1 + g^2 S_2 + g^3 S_3 + \dots \, .
	\end{align} 
	\item $S$ has a block matrix structure. In particular its matrix elements between pairs of states in ${\cal H}_l$ vanish, and it is anti--unitary:
	\begin{align}
		S = \begin{pmatrix}
			0 & -s^\dagger_{hl} \\
			s_{hl} & 0
		\end{pmatrix}\, .
	\end{align}
	\item The canonical transformation induced by $e^S$ yields a block diagonalisation of the full Hamiltonian
	\begin{align}
		e^S\left(H_0+g V\right)e^{-S} = \begin{pmatrix}
			H^{SW}_\text{eff} & 0 \\
			0 & H_{hh}
		\end{pmatrix}\, .
	\end{align}
\end{enumerate}

The definition in \reef{eq:swdef} will automatically produce a Hermitian effective Hamiltonian as $S$ is anti--unitary. Furthermore, the similarity transformation of the Hamiltonian induced by $e^S$ preserves eigenvalues, and taking only the top left block of a block diagonal matrix will not affect eigenvalues either. The SW effective Hamiltonian energy eigenvalues will automatically match their counterparts in the full theory.

By working perturbatively in $g$, the SW Hamiltonian may be determined explicitly. The identity for matrix exponentials $e^ABe^{-A} = B + [A,B]+\dots$ can be used to obtain
\begin{multline}
	e^S\left(H_0+g V\right)e^{-S} = H_0+g\left(V+[S_1,H_0]\right)\\+g^2\left(\frac{1}{2}[S_1,[S_1,H_0]]+[S_1,V]+[S_2,H_0]\right)+\dots\, .
	\label{eq:sworders}
\end{multline}
Working to $O(g)$, the elements of the matrix equation above in the two off--diagonal blocks can be made to vanish by taking
\begin{align}
	[S_1,H_0] =- V_\text{off-diag}\, ,
\end{align}
where $V_\text{off-diag}$ represents the off--diagonal blocks of $V$. This yields the following result for $s_{hl}$ to linear order in $g$
\begin{align}
	s_{hl} = \frac{V_{hl}}{E_h-E_l}\, .
\end{align}
Plugging this result into \reef{eq:sworders}, and taking only the top left matrix block gives the SW Hamiltonian up to second order in the coupling
\begin{align}
	\left(H^{SW}_\text{eff}\right)_{fi} = \delta_{fi}E_i + g V_{fi} + \frac{g^2}{2}\left(\frac{V_{fh}V_{hi}}{E_{fh}}+\frac{V_{fh}V_{hi}}{E_{ih}}\right) + O(g^3)\, ,
\end{align}
where $E_{jk}\equiv E_j-E_k$ and we have also used the fact that the top left block of $[S_2,H_0]$ is zero, which follows from block matrix structure of $S$. A sum over the states of the high energy subspace ${\cal H}_h$ is implied by the repeated index $h$ in the numerators of the $O(g^2)$ terms. If the full theory is a UV regulated and renormalised QFT, the sum over $h$ will yield a finite result.

The higher order contributions to $H^{SW}_\text{eff}$ can be calculated by expanding \reef{eq:sworders} to higher orders in $g$, demanding that the off diagonal blocks of $e^S\left(H_0+g V\right)e^{-S}$ vanish at every order, and then reading off the top left block. At third order, the result is
\begin{align}
	\left(H^{SW}_\text{eff 3}\right)_{fi}=\frac{1}{2}\frac{V_{fh_1}V_{h_1h_2}V_{h_2i}}{E_{h_1f}E_{h_2f}}
	-\frac{1}{2}\frac{V_{fl}V_{lh}V_{hi}}{E_{hf}E_{hl}}+\text{h.c.}
\end{align}
In this instance, all the repeated $h_i$ and $l$ numerator indices are summed over. The $h_i$ label states in ${\cal H}_h$, whereas the $l$ index labels states in ${\cal H}_l$.

The fourth order result is:
\begin{multline}
	\left(H^{SW}_\text{eff 4}\right)_{fi}=-\frac{1}{2}\frac{V_{fh_1}V_{h_1h_2}V_{h_2h_3}V_{h_3i}}{E_{h1f}E_{h2f}E_{h3f}}+\frac{1}{2}\frac{V_{fh_1}V_{h_1h_2}V_{h_2l_1}V_{l_1i}}{E_{h_1i}E_{h_1l_1}E_{h_2l_1}}
	\\+\frac{1}{2}\frac{V_{fh_1}V_{h_1h_2}V_{h_2l_1}V_{l_1i}}{E_{h_1i}E_{h_2i}E_{h_1l_1}}-\frac{1}{2}\frac{V_{fh_1}V_{h_1l_1}V_{l_1l_2}V_{l_2i}}{E_{h_1i}E_{h_1l_1}E_{h_1l_2}}\\
	+\frac{1}{3!}\frac{V_{fh_1}V_{h_1l_1}V_{l_1h_2}V_{h_2i}}{E_{h_1f}E_{h_2f}E_{h_1l_1}}+\frac{2}{3!}\frac{V_{fh_1}V_{h_1l_1}V_{l_1h_2}V_{h_2i}}{E_{h_1f}E_{h_2f}E_{h_2l_1}}+\frac{1}{3!}\frac{V_{fh_1}V_{h_1l_1}V_{l_1h_2}V_{h_2i}}{E_{h_2f}E_{h_1l_1}E_{h_2l_1}}\\
	-\frac{1}{4!}\frac{V_{fh_1}V_{h_1l_1}V_{l_1h_2}V_{h_2i}}{E_{h_1f}E_{h_2l_1}E_{h_1l_1}}-\frac{3}{4!}\frac{V_{fh_1}V_{h_1l_1}V_{l_1h_2}V_{h_2i}}{E_{h_1f}E_{h_1l_1}E_{h_2i}} + \text{h.c.}
\end{multline}
Again, repeated indices $l_i$ and $h_i$ in the numerators indicate sums over all states in the subspaces ${\cal H}_l$ and ${\cal H}_h$ respectively.

\subsection{Connection with Rayleigh--Schr{\"o}dinger Perturbation Theory}
\label{app:rs}

Consider the following operator, acting on the low energy subspace of states ${\cal H}_l$:
\begin{align}
	\matrixel{f}{H^{RS}_\text{eff}}{i} = \frac{\matrixel{f}{\left[\Sigma(H_0+V)\Sigma^\dagger\right]_l}{i}}{\matrixel{f}{\Sigma_l}{f}\matrixel{i}{\Sigma_l}{i}^{-1}}\, .
	\label{eq:RS}
\end{align}
By \reef{eq:fulldiag}, this operator would be a simple diagonal matrix of exact energy eigenvalues. We may therefore regard \reef{eq:RS} as an alternative effective Hamiltonian.

In practice, $\Sigma_l$ and \reef{eq:RS} need to be evaluated using perturbation theory. The diagonal elements of (\ref{eq:RS}) will end up being the Rayleigh--Schr{\"o}dinger (RS) perturbation theory expressions for state energies. Each successive order in RS perturbation theory is suppressed with respect to the last by factors that scale as $V_{ij}/E_{ij}$. For closely separated energy levels $i$ and $j$ in ${\cal H}_l$ and a strong interaction $V$, this factor may be large, in which case truncating the RS series at any fixed order would introduce a large error. For this reason, \reef{eq:RS} would be a poor choice of effective Hamiltonian.

By contrast, elements of the CFHL Hamiltonian in \reef{eq:hl2}, or the SW Hamiltonian in \reef{eq:swdef},  have a much more convergent perturbative series. Successive orders are suppressed by factors that scale as $V_{lh}/(E_l-E_h)$ with $l\in{\cal H}_l$ and $h\in{\cal H}_h$. At least for the states with energies far below the cutoff, the denominator will be large $|E_l-E_h|\gtrsim \D/R$.

\section{Generalisation of the Operator Product Expansion}
\label{app:ope}

In this section, we derive a useful formula for simplifying CFT correlation functions. It can be applied whenever a subset of operators are close together, so that the distances between each of them and a chosen point are all smaller than the distance from that point to any other operator. When only two operators are included in the subset of nearby operators, using this formula is no different from applying the operator product expansion (OPE). However, this formula is more general and can be applied when three or more operators are included in the subset of nearby operators.

Correlation functions in a CFT can be interpreted using radial quantisation (see Ref.~\cite{Simmons-Duffin:2016gjk} for a pedagogical introduction), in which case they act in the order determined by their distances from a given point $y$
\begin{align}
	\L\calO_1(x_1)\dots\calO_n(x_n)\R = \bra{0}\calO_{P(n)}(x_{P(n)})\dots\calO_{P(1)}(x_{P(1)})\ket{0}\,,\label{cfn}\\
	\text{where }|x_{P(1)}-y|\le|x_{P(2)}-y|\le\dots|x_{P(n)}-y|\,,\label{rorder}	
\end{align}
and $P$ is the permutation of indices $1\dots n$ which ensures \reef{rorder}. In general, the ordering of operators in \reef{cfn} depends on the choice made for $y$, the radial quantisation origin.

We can also insert a complete set of states between any pair of operators in \reef{cfn} and leave the correlation function unchanged
\begin{align}
	\L\calO_1(x_1)\dots\calO_n(x_n)\R =\sum_k\bra{0}\calO_{P(n)}\dots\calO_{P(m)}\ket{k}\bra{k}\calO_{P(m-1)}\dots\calO_{P(1)}\ket{0}\,,
\end{align}
where $m$ is an integer in the range $2<m<n$, and we have suppressed the spatial indices of the operators for brevity. Using the state operator correspondence, a set of states which completely spans the CFT Hilbert space can be constructed by acting on the vacuum with all possible local operators
\begin{align}
	\ket{k}\rightarrow \calO(y)\ket{0}\,,\qquad\qquad \bra{k}\rightarrow\lim\limits_{s\rightarrow\infty}|s|^{2\Delta_{\calO}}\bra{0}\calO(s)\equiv\bra{0}\calO(\infty)\,,
\end{align}
so that
\begin{align}
	\L\calO_1(x_1)\dots\calO_n(x_n)\R&=\sum_{\calO}\bra{0}\calO_{P(n)}\dots\calO_{P(m)}\calO(y)\ket{0}\bra{0}\calO(\infty)\calO_{P(m-1)}\dots\calO_{P(1)}\ket{0}\,,\\
	&=\sum_{\calO}\L\calO_{P(n)}\dots\calO_{P(m)}\calO(y)\R\L\calO(\infty)\calO_{P(m-1)}\dots\calO_{P(1)}\R\,.\label{genresult}
\end{align}
The sum runs over all CFT operators, including both primaries and descendants. The formula above is a general result, valid for any CFT correlation function and any choice of radial quantisation origin $y$.

When $m=3$, using the formula \reef{genresult} is equivalent to using the OPE. To demonstrate this explicitly, we take $y=x_2$ and make the identification
\begin{align}
	C_{12j}(x_{12},\partial_2)\calO_j(x_2)=\sum_{\text{desc. of }j}\calO(x_2)\L\calO(\infty)\calO_1(x_1)\calO_2(x_2)\R\,,
\end{align}
where the sum runs over primary operator $\calO_j$ and all of its descendants. Then \reef{genresult} gives
\begin{align}
	\L\calO_1(x_1)\dots\calO_n(x_n)\R&=\sum_jC_{12j}(x_{12},\partial_2)\L\calO_j(x_2)\calO_3(x_3)\dots\calO_n(x_n)\R\,,
\end{align}
provided that $|x_{12}|<|x_{23}|,\dots|x_{2n}|$. We see that we have the same result you would get by using the OPE for operators $\calO_1$ and $\calO_2$.

For $m=4$ and $n=5$, we derive another useful result, which we use to simplify the computation of the third order effective hamiltonian in Section~\ref{sec:thirdo}. If we consider the matrix element below and radially order it, picking $y=1$ as our radial quantisation origin, we find
\begin{align}
	\matrixel{f}{\phi_\Delta(1)\phi_\Delta(x_2)\phi_\Delta(x_1)}{i} =\L\calO_f(\infty)\calO_i(0)\phi_\Delta(x_2)\phi_\Delta(x_1)\phi_\Delta(1)\R\,,
\end{align}
where we have assumed that $|1-x_1|$, $|1-x_2|<1$ when radially ordering. Applying \reef{genresult} then yields
\begin{align}
	\matrixel{f}{\phi_\Delta(1)\phi_\Delta(x_2)\phi_\Delta(x_1)}{i} &=\sum_{\calO}\L\calO_f(\infty)\calO_i(0)\calO(1)\R\L\calO(\infty)\phi_\Delta(x_2)\phi_\Delta(x_1)\phi_\Delta(1)\R\,,\\
	&=\sum_{\calO}\matrixel{f}{\calO(1)}{i}\L\calO(\infty)\phi_\Delta(1)\phi_\Delta(x_2)\phi_\Delta(x_1)\R\,,
\end{align}
as shown in \reef{factor}.

\section{Higher order corrections and $\phi^4$ theory}
\label{appv6}

In this section we would like to explain  how to apply the construction that we have introduced in the main text to the $\phi^4$ perturbation of the  free massive scalar theory. 
The derivations that we perform next apply more generally and have a structural similarity with Conformal Perturbation Theory. We elucidate this point at the end of the section.

\subsection{Fourth order corrections}

In section~\ref{sec:cutoffdep} we explained how locality is restored to $O(V^4)$ in the context of conformal perturbation theory. 
Specifically we showed that $K_2$ cancels unphysical UV divergences in the Casimir energy at $O(V^4)$.
The purpose of this section is to derive the analogous result in the context of the  $\phi^4$ perturbation of free massive theory.

Recall that 
\be
\Eps^{(4)}_0=  -\frac{  V_{0k_1 }V_{k_1 k_2}V_{k_2  k_3} V_{k_3  0}  }{E_{k_1}E_{k_2}E_{k_3} } - \Eps^{(0)}_2  \frac{V_{0k}V_{k0}}{E_k^2} \, .
\label{eps4app}
\ee
In Ref.~\cite{Elias-Miro:2020qwz} it was shown that the following  contributions  to the first term in \reef{eps4app} grow the fastest in the limit $E_T \rightarrow \infty$
\be
-\dVdos-\dVtres  = -  \int_{4m}^{ E_1+E_2\leq E_T  }  \left(   
\frac{1}{E_1}\frac{1}{E_1+E_2}  \frac{1}{E_2}  +\frac{1}{E_1}\frac{1}{E_1+E_2}  \frac{1}{E_1} \right) \  d\widetilde E_1 d\widetilde E_2  \,  ,    \label{fock1}
\ee
where have defined $d\widetilde E_i\equiv dE_i/(2\pi) \Phi_4(E_i)$, and where where $\Phi_4(E_i)$ is the four-particle phase space.~\footnote{We are omitting an inconsequential overall normalisation in \reef{fock1} and \reef{fock2}, given by $((gL)^2/24)^2$~\cite{Elias-Miro:2020qwz}.}
In particular, because $\Phi_4(x)\sim x$ in $d=2+1$ spacetime dimensions, these diagrams diverge as $E_T\rightarrow \infty$, for this spacetime dimension. 
The notation $ \int_{4m}^{x\leq E_T  } \equiv  \int_{4m}^{\infty } \ \theta(x- E_T)  $, where $\th$ is the Heaviside step function. 
The second term in  \reef{eps4app} is given by
\be
 - \Eps^{(0)}_2 \times   \frac{V_{0k}V_{k0}}{E_k^2} = \int_{4m}^{ E_T} \frac{1}{E_1^2} \frac{1}{E_2^2}    \  d\widetilde E_1 d\widetilde E_2 \, .  \label{fock2}
\ee
Because in Hamiltonian Truncation the states have energy  bounded by $E_T$, in \reef{fock1} we have indicated  by dashed vertical lines cuts through the diagrams indicating the propagating energy between any to consecutive vertices. 
These are \emph{old fashioned perturbation theory} diagrams, where time flows horizontally, lines denote on-shell particles and the order of the vertices  in the time direction matters.  

In $\phi^4$ theory in $d=2+1$ dimensions we do not expect any UV divergence in the vacuum energy at $O(V^4)$. 
However,  when truncating the theory   $H_0+ V$,  with $V= \int d^2x \phi^4$,
  \reef{fock1} and \reef{fock2} do not cancel~\cite{Elias-Miro:2020qwz}
\be
\text{\reef{fock1}}+\text{\reef{fock2}}\sim E_T- 8 \log (E_T/m ) + O(E_T^0) \, . 
\ee
 because  of truncation on the maximal propagating energy.
 It turns out that this new type of divergences appear at all orders in perturbation theory. 
 
 In this work we have argued that the way to deal with these effects is to perform usual local-regulator renormalisation, and then computed an Effective Hamiltonian to the same order as the needed local conterterms. 
 In this way we are are able derive an effective truncated Hamiltonian
 \be
 H_\text{eff} = H_0+V + K \label{heffapp}
 \ee
 where the dependence on the local regulator has been removed. 
 Carrying out this procedure  for the vacuum two-point divergence, we are left with 
 \be
 (K_2)_{ij}=  k(E_1) \delta_{ij}+ \dots \quad \text{where} \quad  k(E_1)=    \int_{4m}^{E_1+E_2\leq E_T} \frac{d\widetilde{E}_2 }{E_2}  
 \label{kapp} \ee
 where the dots $\cdots$ denote contributions not proportional to the identity operator. 
 
 In order to re-calculate the two-point function contributions to $\Eps^{(4)}_0$, first we note that  \reef{kapp} sets to zero $\Eps^{(2)}_0$. 
 Therefore we do not have the contribution in \reef{fock2} in the $(H_0+V)_{ij}+ k(E_i)\delta_{ij}$ theory. 
 Next we need to take into account the diagrams with a single insertion of $k$. The only contribution  is given by
\be
\ctvdos
 =  -  \int_{4m}^{E_1 \leq  E_T }   
\frac{1}{E_1^2} k(E_1) \  d\widetilde E_1\, . \label{canc0}
\ee
The operator $K_2$ is a non-local interaction and therefore we do not represent as a vertex in a diagram. 
It is instead represented by  acting on all particles at a given time-slice, irrespective of the distance among the particles. 
Next note that  \reef{canc0} nicely cancels  \reef{fock1}, thus $\Eps^{(4)}_0$ is finite in this theory!

We remark that we have only added a second orrder operator $K_2$ to the Hamiltonian, and nonetheless it has cured an inconsistency of \emph{raw} Hamiltonian Truncation  at order $O(V^4)$.
Next we present an explicit check that this consistency continues to hold at higher orders.

\subsection{Sixth order corrections}

Here we discuss the most divergent contributions to the vacuum energy at sixth order, that arise from diagrams involving    two-point functions only.
These arise from the following term
\be
\Eps^{(6)}_0=  \frac{  V_{0k_1 }V_{k_1 k_2}V_{k_2  k_3} V_{k_3 k_4} V_{k_4 k_5} V_{k_5 0}  }{E_{k_1}E_{k_2}E_{k_3}E_{k_4}E_{k_5}  } + \cdots  \, .  \label{eps6}
\ee
The $\cdots$ involve terms that are unimportant for our current purposes. 
There are many diagrams with vacuum two-point functions only that contribute to the first term in \reef{eps6}.
It is useful to divide them into two groups. 
The first group consists of diagrams that  contain  cuts involving two vacuum bubbles at most:
\bea
 \dun &  = \int_{4m}^{\substack{ E_1+E_2\leq E_T \\  E_2+E_3\leq E_T} }    
\frac{1}{E_1}\frac{1}{E_1+E_2}  \frac{1}{E_2}\frac{1}{E_2+E_3}\frac{1}{E_3}  \  d\widetilde E_1 d\widetilde E_2 d\widetilde E_3  \label{b9app}  \,  , \\
\ddos &= \int_{4m}^{\substack{ E_1+E_3\leq E_T \\  E_2+E_3\leq E_T} } 
\frac{1}{E_1}\frac{1}{E_1+E_2}  \frac{1}{E_1}\frac{1}{E_1+E_3}\frac{1}{E_3}  \  d\widetilde E_1 d\widetilde E_2 d\widetilde E_3   \, , \\ 
\dtres &= \int_{4m}^{\substack{ E_1+E_3\leq E_T \\  E_1+E_2\leq E_T} } 
\frac{1}{E_2}\frac{1}{E_1+E_2}  \frac{1}{E_1}\frac{1}{E_1+E_3}\frac{1}{E_1}  \  d\widetilde E_1 d\widetilde E_2 d\widetilde E_3   \, , \\ 
\dquatre &=  \int_{4m}^{\substack{ E_1+E_3\leq E_T \\  E_2+E_3\leq E_T} }   
\frac{1}{E_1}\frac{1}{E_1+E_2}  \frac{1}{E_1}\frac{1}{E_1+E_3}\frac{1}{E_1} \  d\widetilde E_1 d\widetilde E_2 d\widetilde E_3    \,   .
\eea
Adding these four diagrams above we are lead to 
\be
 \int_{4m}^{\substack{ E_1+E_3\leq E_T \\  E_2+E_3\leq E_T} }   
 \frac{1}{E_1^3}\frac{1}{E_2}\frac{1}{E_3} \  d\widetilde E_1 d\widetilde E_2 d\widetilde E_3   \label{canc1}
\ee
The second group   involves  three bubbles overlapping at the same moment in time:
\bea
\dcinc & = \int_{4m}^{ E_1+E_2+E_3\leq E_T }  
\frac{1}{E_1}\frac{1}{E_1+E_2}  \frac{1}{E_1+E_2+E_3}\frac{1}{E_2+E_3}\frac{1}{E_3}   \ d\widetilde E_1d\widetilde E_2d\widetilde E_3   \,  , \label{d5}\\
\dsis & =   \int_{4m}^{ E_1+E_2+E_3\leq E_T }  
\frac{1}{E_1}\frac{1}{E_1+E_2}  \frac{1}{E_1+E_2+E_3}    \frac{1}{E_1+E_3}\frac{1}{E_3}  \ d\widetilde E_1d\widetilde E_2d\widetilde E_3  \,   ,   \label{d6}\\
 \dsishc&= \int_{4m}^{ E_1+E_2+E_3\leq E_T }    \frac{1}{E_1}\frac{1}{E_1+E_2}  \frac{1}{E_1+E_2+E_3}    \frac{1}{E_2+E_3}\frac{1}{E_2}  \ d\widetilde E_1d\widetilde E_2d\widetilde E_3  \label{d6hc}\\
\dset &= \int_{4m}^{ E_1+E_2+E_3\leq E_T }  
\frac{1}{E_1}\frac{1}{E_1+E_2}  \frac{1}{E_1+E_2+E_3}\frac{1}{E_1+E_2}\frac{1}{E_2}  \ d\widetilde E_1d\widetilde E_2d\widetilde E_3   \,  ,  \label{d7} \\
\dvuit & = \int_{4m}^{ E_1+E_2+E_3\leq E_T }  
\frac{1}{E_1}\frac{1}{E_1+E_2}  \frac{1}{E_1+E_2+E_3}\frac{1}{E_1+E_3}\frac{1}{E_1}  \ d\widetilde E_1d\widetilde E_2d\widetilde E_3 \, ,    \label{d8}\\
\dnou & =  \int_{4m}^{ E_1+E_2+E_3\leq E_T }  
\frac{1}{E_1}\frac{1}{E_1+E_2}  \frac{1}{E_1+E_2+E_3}\frac{1}{E_1+E_2}\frac{1}{E_1}    \ d\widetilde E_1d\widetilde E_2d\widetilde E_3 \label{d9} \, .
\eea
Adding all these diagrams we are lead to 
\be
\int_{4m}^{ E_1+E_2+E_3\leq E_T }    \frac{1}{E_1^2}\frac{1}{E_2}\frac{1}{E_3}\frac{1}{E_1+E_2}  \ d\widetilde E_1d\widetilde E_2d\widetilde E_3  \, .  \label{canc2}
\ee

Next we need to consider diagrams involving $K_2$. 
Again it is useful to separate diagrams into two groups. 
The first group contain a cut with at most two two-point vacuum bubbles:
\bea
\ctun + \ctdos & = - 2 \int_{4m}^{E_1+E_2\leq E_T}  \frac{1}{E_1}k(E_1)\frac{1}{E_1} \frac{1}{E_1+E_2} \frac{1}{E_2}  \  d\widetilde E_1 d\widetilde E_2  \, , \label{ct1} \\ 
\cttres +\ctquatre&= -2 \int_{4m}^{E_1+E_2\leq E_T}  \frac{1}{E_1}k(E_1)\frac{1}{E_1} \frac{1}{E_1+E_2} \frac{1}{E_1}  \  d\widetilde E_1 d\widetilde E_2   \,   ,  \label{ct2} \\
\ctcinc&=+ \int_{4m}^{E_1\leq E_T}  \frac{1}{E_1}k(E_1)\frac{1}{E_1} k (E_1)  \frac{1}{E_1}    \  d\widetilde E_1   \,   .  \label{ct3}
\eea
Adding the last three equations we get an exact cancelation against \reef{canc1}.
The remaining diagrams involving  $K_2$ are given by 
\be
 \ctsis+ \ctset =  -   \int_{4m}^{ E_1+E_2\leq E_T }    
\frac{1}{E_1^2}\frac{1}{E_2}\frac{1}{E_1+E_2}  k(E_1+E_2)  \  d\widetilde E_1 d\widetilde E_2 d\widetilde E_3  \, .   \label{cttot}
\ee
The last equation exactly cancels \reef{canc2}.
This calculation explicitly shows the restoration of locality at $O(V^6)$!

\subsection{Generalisations}

In this appendix we have shown that   $K$ restores locality of the Hamiltonian Truncation computations. 
We did so for the most UV divergent diagrams only. These consist in diagrams containing the maximum  possible number of  disconnected two-point functions. 

The computations here presented can be easily generalised because the   particularities of the specific  theory are hidden in the measure of the integrals $d \widetilde E_i$.
Therefore the analysis is  easily extended to any spacetime dimension or theory by picking the correct phase-space functions. 
Importantly the structure of cancelations between diagrams involving  $K$ and  diagrams involving only $H_0+V$ will remain true. 
It also generalises to Conformal Perturbation Theory~\footnote{Of course in the far UV it should not matter whether the calculation is organised as a perturbation of a  massive Fock space or as a Conformal Perturbation Theory.} where 
\be
\int_{4m}^{\infty} d \widetilde E_i  \longrightarrow    \int_0^\infty d E_i   \sum_{n=0}^{\infty } u_n^{2\Delta} 	\, \delta( E_i - (2n+\Delta)/R)
\ee
Then each diagram in 
the previous section  has an analog as an integral of an $n$-point function in the TCSA cylinder with a particular time ordering. 
For instance
\be
\dVdos  \quad \longrightarrow \quad 
 \begin{minipage}[h]{0.1\linewidth}
  \includegraphics[width=1.9cm]{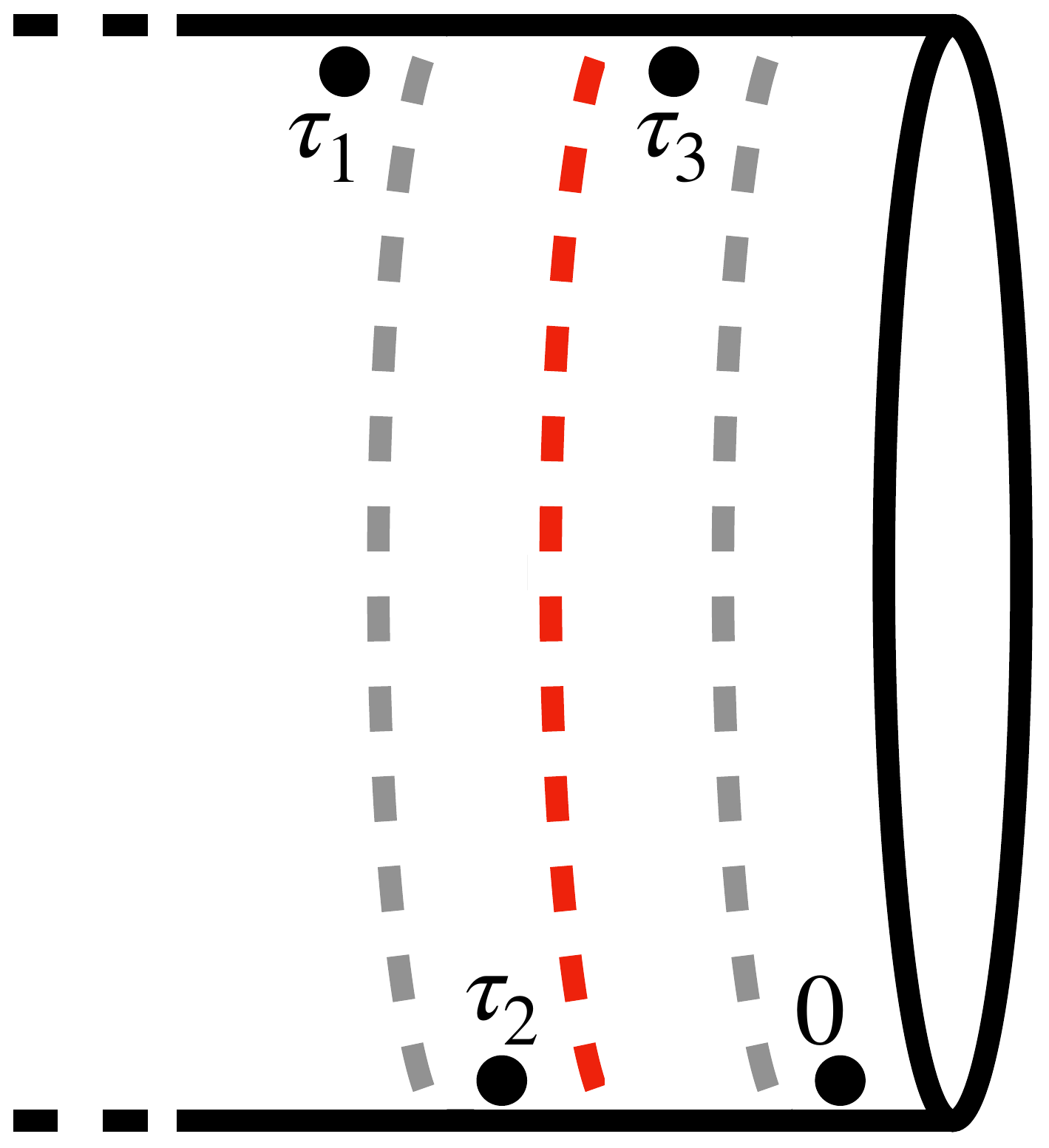} 
  \end{minipage}
\ee
which is given by the following integral~\cite{EliasMiro:2021aof}
\be
\int \prod_{i=1}^3 d^{d-1}x_i \sum_{k,k^\prime}^{\Delta_k + \Delta_{k^\prime} \leq \Delta_T} \frac{1}{\Delta_{k}\Delta_{k^\prime}(\Delta_k + \Delta_{k^\prime})}  
\langle 0  | \phi_{0,\vec n} |k^\prime \rangle \langle k^\prime  | \phi_{0,\vec{x}_2} |0 \rangle
 \langle 0  | \phi_{0,\vec{x}_3} |k^\prime \rangle \langle k^\prime  | \phi_{0,\vec{x}_1} |0 \rangle \, ; 
\ee
or, for instance
\be
\dVtres   \quad \longrightarrow \quad 
 \begin{minipage}[h]{0.1\linewidth}
  \includegraphics[width=1.9cm]{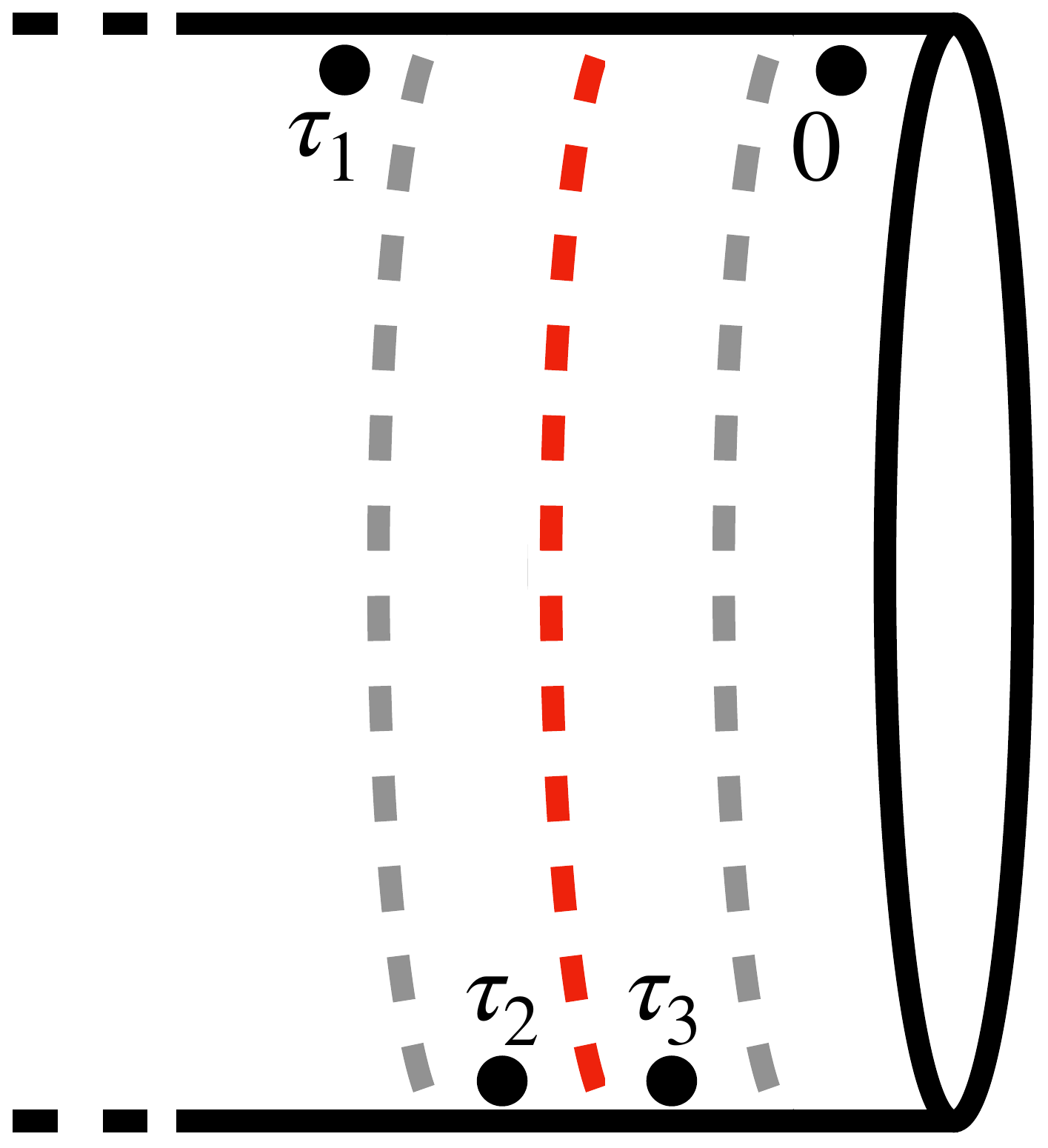} 
  \end{minipage}
\ee
which is instead  given by
\be
\int \prod_{i=1}^3 d^{d-1}x_i \sum_{k,k^\prime}^{\Delta_k + \Delta_{k^\prime} \leq \Delta_T} \frac{1}{\Delta_{k^\prime}^2(\Delta_k + \Delta_{k^\prime})}  
\langle 0  | \phi_{0,\vec n} |k^\prime \rangle \langle k^\prime  | \phi_{0,\vec{x}_1} |0 \rangle
 \langle 0  | \phi_{0,\vec{x}_3} |k^\prime \rangle \langle k^\prime  | \phi_{0,\vec{x}_2} |0 \rangle \, .
\ee

\section{Useful Formulae For Evaluating $K_3$}
\label{sec:formulae}

In this section, we evaluate the integral in \reef{xlim} to obtain an expression for it in the form of a finite sum of terms, depending on elementary functions. This function can be evaluated numerically more easily than the integral.

To begin, we express the integrand as a linear combination of Gegenbauer polynomials $C^{\alpha}_l(\cos\theta)$ with $\alpha=(d-2)/2$
\begin{align}
	C_l^\Delta(\cos\theta)=\sum_{k=0}^l\calA_{l,k}\;C_k^{\frac{d-2}{2}}(\cos\theta)\,,\label{gegbasis}
\end{align}
where the expansion coefficients are found to be
\begin{multline}
	\calA_{l,k}=\frac{2^{1-2k-2\Delta}\sqrt{\pi}\,\Gamma\left(\frac{d-2}{2}\right)\Gamma\left(k+l+2\Delta\right)}{\Gamma\left(\Delta\right)\Gamma\left(l-k+1\right)\Gamma\left(k+\Delta+\frac{1}{2}\right)\Gamma\left(k+\frac{d}{2}-1\right)}\\\times{}_3F_2\left(k+\frac{d-1}{2},k-l,k+l+2 \Delta;2 k+d-1,k+\Delta+\frac{1}{2};1\right)\,,
\end{multline}
after using results from Appendix G of \cite{Anand:2020gnn}. Since Gegenbauer polynomials $C^{\alpha}_l(\cos\theta)$ are odd functions in $\cos\theta$ when $l$ is odd, and even functions when $l$ is even, the coefficients $\calA_{l,k}$ will vanish unless $l$ and $k$ are both even or both odd. These polynomials are orthogonal with respect to the measure for spherical coordinates in $d-1$ dimensions, and therefore form a more convenient basis of functions
\begin{align}
	\int_{S_{d-1}}d^{d-1}x\;C^{(d-2)/2}_{n}(\cos\theta)\,C^{(d-2)/2}_{m}(\cos\theta)=\delta_{nm}\frac{2^{3-d}\pi\Gamma(n+d-2)}{n!\left(n+\frac{d-2}{2}\right)\left[\Gamma\left(\frac{d-2}{2}\right)\right]^2}S_{d-2}\,.\label{ortho}
\end{align}
The measure above can be written explicitly as $d^{d-1}x=\left(\sin\theta\right)^{d-2}d\theta\left(\sin\phi\right)^{d-3}d\phi\dots$ in terms of the $d-1$ angular coordinates.

To do the integral in \reef{xlim}, we orient the two sets of spherical coordinates so that $\cos\gamma=\cos\theta_1\cos\theta_2+\sin\theta_1\sin\theta_2\cos\phi_1$. We then use the Gegenbauer polynomial addition formula \cite{Bateman:100233} shown below
\begin{multline}
	C^{\frac{d-2}{2}}_n(\cos\gamma)=\sum_{m=0}^nB_{n,m}\left(\sin\theta_1\right)^mC_{n-m}^{m+\frac{d-2}{2}}(\cos\theta_1)\left(\sin\theta_2\right)^mC_{n-m}^{m+\frac{d-2}{2}}(\cos\theta_2)\,C_k^{\frac{d-3}{2}}(\cos\phi_1)\,.\label{gegadd}
\end{multline}
The only term in \reef{gegadd} that survives after integrating over $\phi_1$
is the $m=0$ term, as a result of the orthogonality relation \reef{ortho} and $1=C_0^{\frac{d-3}{2}}(\cos\phi)$. The only coefficients from \reef{gegadd} that we need are
\begin{align}
	B_{n,0}=\frac{n!\Gamma(d-2)}{\Gamma(d-2+n)}\,.
\end{align}

Plugging the series expansions \reef{gegbasis} and then \reef{gegadd} into \reef{xlim}, and integrating using \reef{ortho} yields
\begin{align}
	X_{n_1,n_2,n_3}=\sum_{k=0}^{\text{Min}(n_1,n_2,n_3)}\frac{\Gamma(k+d-2)}{k!\left[\Gamma(d-2)\right]^{-1}}\left(\frac{2^{3-d}\pi\,S_{d-2}}{\left(k+\frac{d-2}{2}\right)\left[\Gamma\left(\frac{d-2}{2}\right)\right]^2}\right)^2\calA_{n_1,k}\,\calA_{n_2,k}\,\calA_{n_3,k}\,,
\end{align}
a convenient expression for numerical evaluation. This function is manifestly symmetric in its $n_i$ indices. It is also nonzero only when all $n_i$ are even, or all of them are odd. This follows because the $\calA_{n,k}$ vanish unless its indices are both even, or both odd.


\small

\bibliography{biblio}
\bibliographystyle{utphys}

\end{document}